Unveiling the critical factors in crystal structure graph representation: a comparative analysis using streamlined MLPSets frameworks

Hongwei Du[1,2,3], Hong Wang*[1,2,3]

1 School of Materials Science and Engineering, Shanghai Jiao Tong University, Shanghai 200240, China.

2 Zhangjiang Institute for Advanced Study, Shanghai Jiao Tong University, Shanghai 201203, China.

3 Materials Genome Initiative Center, Shanghai Jiao Tong University, Shanghai 200240, China.

Corresponding authors
Correspondence to: Hong Wang, hongwang2@sjtu.edu.cn (Hong Wang).

# Abstract

Graph Neural Networks (GNNs) have rapidly advanced in the fields of materials science and chemistry, becoming essential tools for predicting material properties and developing machine learning potential models. The core of GNNs lies in the comprehensive representation of crystal or molecular structures through five key dimensions: elemental information, geometric topological information, electronic interaction information, symmetry information, and long-range interaction information. By integrating these dimensions, GNNs can more comprehensively describe material characteristics. For example, MatterSim uses pre-trained elemental embeddings to represent chemical information, while DPA-2 enhances elemental information representation through pre-trained knowledge from the periodic table. In terms of geometric topological information, methods have evolved from early CGCNN to ALIGNN and M3GNet, showcasing continuous advancements in this dimension. Despite significant progress, existing models still fall short in providing a complete representation of crystal structures, particularly in the representation of electronic interactions, lacking strategies similar to "pseudopotentials" in Density Functional Theory (DFT) calculations. To delve into the critical factors affecting the performance of crystal structure graph representations, we designed and compared two types of crystal structure graph representation strategies: physics-based site feature calculators and data-driven structure graph representation strategies. Physics-based site calculator models such as AGNIFingerprints, OPSiteFingerprint, and CrystalNNFingerprint each have unique strengths in representing local geometric topology and chemical environments but are limited in describing symmetry and long-range interaction information. Data-driven graph representation strategies, such as MEGNet, M3GNet, and the CHGNet series, perform better in multi-dimensional representation of crystal structures. Specifically, by incorporating an electronic structure generation model (variational autoencoders VAE), which compresses Kohn–Sham (KS) wave functions into a low-dimensional latent space using an encoder and reconstructs them using a decoder, these models efficiently represent electronic structures. Combined with GNN network design and multi-task material property label feedback learning, they significantly enhance the comprehensive representation of crystal structures. Our study demonstrates that the data-driven CHGNet-V1 and V2 structural graph representation strategies exhibit superior performance and faster convergence in multiple benchmark tasks, as well as significant robustness and consistency in extrapolation tasks. We applied the CHGNet-V1 graph representation strategy to the advanced DenseGNN model. Extensive testing on 35 different datasets from Matbench and JARVIS-DFT showed that the DenseGNN-CHGNet model significantly outperformed current state-of-the-art models. Additionally, DenseGNN-CHGNet provides high-fidelity prediction datasets that closely match DFT accuracy, facilitating the application of transfer learning strategies in materials research. Finally, by pre-training DenseGNN-CHGNet on the Matbench band gap dataset and fine-tuning it with disordered material data, we significantly reduced the mean absolute error (MAE) in predicting the band gaps of complex disordered materials such as $Al_xGa_{1-x}N$ and $Mg_xZn_{1-x}O$. This underscores the importance of data-driven graph representation strategies in enhancing extrapolation performance and their potential applications in the discovery of new materials, providing new insights into the comprehensive representation of crystal structures.

## Introduction

Graph Neural Networks (GNNs) have rapidly advanced in the fields of materials science and chemistry, becoming a critical tool for predicting material properties and developing machine learning potential models[1-9]. The core of GNNs lies in the comprehensive representation of crystal or molecular structures to accurately capture the physical relationships between structure, composition, and properties. This representation primarily encompasses five key dimensions: elemental information, geometric topological information, electronic interaction information, symmetry information, and long-range interaction information. By integrating these dimensions, GNNs can provide a more holistic description of material characteristics. For example, MatterSim[10] employs pre-trained elemental embeddings to capture chemical information, while DPA-2[11] enhances elemental information expression through pre-trained knowledge from the periodic table. In terms of geometric topological information, the representation methods have evolved from the early binary system of CGCNN to iCGCNN's Voronoi cell-based approach[12, 13], ALIGNN's[14] multi-layer nested dihedral-angle-edge subgraph representation, and M3GNet's[2] multi-body interaction module, which describes the neighbor geometric topological information of elemental sites. These methods have become increasingly sophisticated and complex. However, the highly nonlinear and complex nature of electronic structure makes it difficult to represent. In particular, the high dimensionality of Kohn-Sham(KS)[15] wave functions and their sensitivity to crystal configurations and non-local correlations pose significant challenges for directly representing electron interactions. Density Functional Theory (DFT)[15] is the most commonly used method for studying electronic structures of materials, but there has been a lack of strategies akin to "pseudopotentials" in DFT for representing electronic interactions in crystal structures. Therefore, indirect methods, such as multi-task pre-training and introducing features like electronegativity, oxidation state, and magnetic moments, are employed to enhance the representation of electronic interactions. Symmetry information and long-range interaction information are realized through equivariant GNN components and message-passing frameworks, as seen in models like EquiformerV2 and MACE[3, 8].

Despite significant progress in the application of GNNs in materials science, there are still notable deficiencies in the representation of electronic interactions. Specifically, existing structural graph representations lack strategies similar to the "pseudopotential" concept in DFT calculations, making it difficult to effectively describe electronic interactions. Additionally, there is a lack of systematic comparative studies to evaluate the effectiveness of different crystal structure graph representation strategies. To explore the key factors affecting the performance of crystal structure graph representations, we designed and compared two types of crystal structure graph representation strategies: one based on physics-driven site feature calculators and the other based on data-driven pre-trained GNN models. Comparative studies of both types of strategies demonstrated that the description of geometric and electronic interactions in crystal structure graphs significantly influences the performance of material property prediction models.

The first type of crystal graph representation strategy is based on three physically motivated site calculators: AGNIFingerprints[16], OPSiteFingerprint[17], and CrystalNNFingerprint[17, 18]. Each of these models has unique strengths in describing the local geometry and chemical environment of crystal structures. AGNIFingerprints captures local geometric features by calculating atom-atom distances with Gaussian-weighted integrals, excelling in spatial arrangement and geometric shape description but lacking in electronic interaction representation. OPSiteFingerprint focuses on describing atomic coordination geometry and symmetry features through structural order parameters, but still lacking in electronic interaction representation. CrystalNNFingerprint integrates structural order parameters and chemical information (such as differences in electron affinity, electronegativity, etc.), enhancing the description of local electronic interactions at atomic sites and performing better in predicting properties related to complex chemical interactions, emphasizing the importance of considering both geometric topology and electronic interaction information. However, these physics-based site calculators are unable to effectively represent symmetry information, electronic and long-range interactions. The second type of crystal graph representation strategy is based on data-driven pre-trained GNN models (MEGNet[19], M3GNet[2], CHGNet-NO-VAE, CHGNet-V1, and CHGNet-V2), all of which were pre-trained on the Materials Project Trajectory Dataset[20]. This dataset contains approximately 1.58 million atomic configurations and dynamic evolution information for 145,000 inorganic materials, providing rich training data for the models. Among these five models, MEGNet can only capture basic crystal structure features. M3GNet and CHGNet-NO-VAE, while capable of representing geometric, symmetry, and long-range interaction information to some extent, are notably limited in their ability to represent electronic interaction. CHGNet-V1 leverages a Variational Autoencoder (VAE)[21, 22] to compress KS wave functions into a low-dimensional latent space using an encoder and then reconstructs them using a decoder, thereby efficiently representing the electronic structure. It combines this with the multi-body interaction modules of CHGNet[1] and label feedback from the training dataset to achieve a comprehensive representation of both geometric and electronic interactions in crystal structures. CHGNet-V2 further enhances the representation by incorporating multi-task label feedback learning with the VAE model, generating a more complete representation of both geometric and electronic interactions in crystal structures.

Based on the above two types of structural graph representation strategies, we introduced the Streamlined MLPSets framework, aiming to minimize the interference of GNN operations such as message passing and graph convolution, thereby simplifying the network design to clearly compare different crystal structure graph representation strategies and identify which strategy performs better. Through comparative studies of these two types of structural graph representation strategies, we found that data-driven structure graph representation strategies (such as CHGNet-V1 and V2) in the MLPSets framework achieved performance comparable to current state-of-the-art reference models (e.g., coGN)[23] on multiple benchmarks (13 datasets from Matbench and JARVIS-DFT)[24, 25] but with simpler architectures and fewer trainable parameters. Additionally, the MLPSets-CHGNet-V1 model not only showed higher performance in benchmark tasks but also converged faster compared to advanced reference models, highlighting the advantage of data-driven structure graph representation strategies in improving model prediction efficiency. In extrapolation tasks, such as predicting shear modulus (log(GVRH)) calculated by DFT, MLPSets-CHGNet-V1, despite its simpler architecture and fewer parameters, achieved significantly higher extrapolation accuracy. Additional tests on other material property datasets further confirmed its robustness and consistency in various material property prediction tasks.

Based on these research findings, we applied the CHGNet-V1 graph representation strategy to three specific application scenarios. First, we applied the CHGNet-V1 graph representation strategy to our previously proposed DenseGNN model[26], resulting in DenseGNN-CHGNet, which showed significant performance improvements on 35 datasets from Jarvis-DFT and Matbench, further demonstrating the superiority of data-driven crystal structure graph representation strategies. Second, DenseGNN-CHGNet can provide high-fidelity datasets with near-DFT calculation accuracy. By comparing the impact of different fidelity training data provided by different GNN models on the transfer performance of MLPSets-CHGNet models in formation energy and band gap experimental datasets, we highlighted the key role of the data-driven CHGNet-V1 crystal structure representation strategy. Finally, based on the DenseGNN-CHGNet model, we used a knowledge transfer strategy, first training on the band gap dataset of Matbench and then fine-tuning on the disordered experimental band gap dataset, not only reducing the Mean Absolute Error (MAE) but also achieving performance comparable to the multi-fidelity MEGNet-4fi model, confirming the effectiveness of the data-driven CHGNet-V1 graph representation strategy in enhancing the description of electronic interactions.

In summary, the comparison of these two types of structural graph representation strategies shows that the description of geometric and electronic interactions in crystal structure graphs significantly affects the performance of material property prediction models. While traditional physics-based site feature calculators have certain advantages in geometric feature description, they are notably deficient in electronic interaction representation. In contrast, data-driven pre-trained GNN models combined with VAE-based graph representation strategies can more comprehensively capture this information. Specifically, the CHGNet-V1 and V2, by integrating electronic structure generation models (VAE), multi-body interaction modules, and label feedback from training datasets, provide new insights into achieving a comprehensive representation of crystal structures.

# Results

## Compare crystal structure graph representation strategies via a streamlined MLPSets frameworks

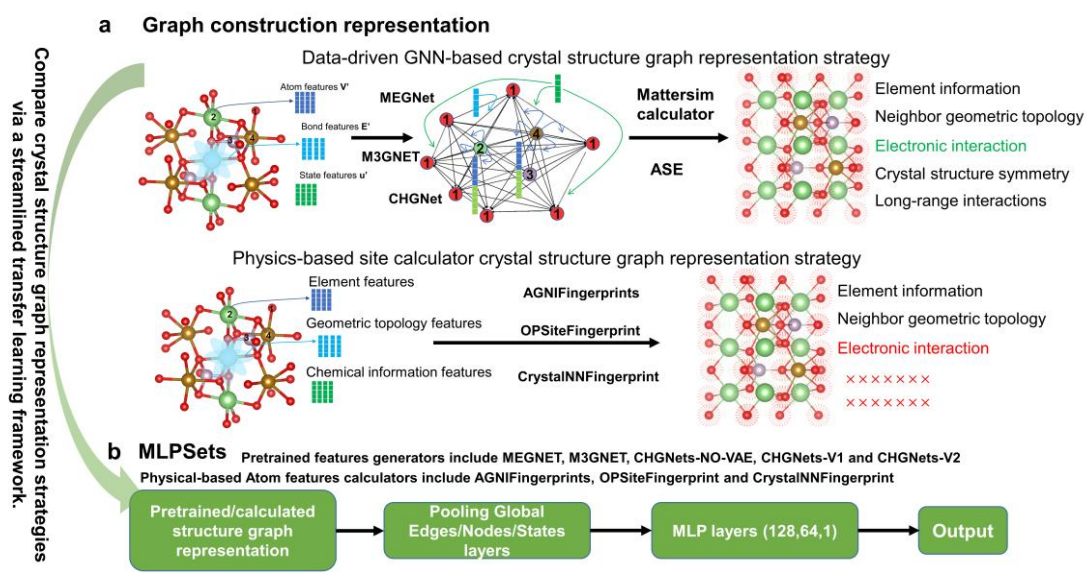

Figure 1 presents a comparison of different crystal structure graph representation strategies based on the streamlined MLPSets framework. In Figure 1a, two approaches for representing crystal structures are compared: one is a data-driven method, which can encode geometric information, electronic interactions, symmetry, and long-range interactions, but its performance heavily relies on the design of the GNN model and the dataset; the other is a physics-based site feature calculator, which excels at capturing local geometric and some chemical environment features but falls short in describing symmetry, electronic and long-range interactions, as indicated by the red X marks in the figure. Figure 1b illustrates the workflow of the MLPSets framework: starting from various types of crystal structure graph representations, it first transforms these into graph-level vectors through an aggregation layer, then processes them using a Multi-Layer Perceptron (MLP), and finally generates structure-level feature vectors via a readout function to predict material properties.

Despite significant advancements in the application of GNNs in materials science, there remains a notable deficiency in the representation of electronic interactions. Current models lack a strategy analogous to the "pseudopotential" concept in DFT calculations, which is essential for effectively describing electronic interactions. Additionally, there is a lack of systematic comparative studies to evaluate the effectiveness of different crystal structure graph representation strategies. To explore the key factors influencing the performance of crystal structure graph representations, Figure 1 illustrates a comparison of crystal structure graph representation strategies through a streamlined MLPSets framework. Figure 1a presents two categories of crystal structure graph representation strategies: one based on data-driven structure graph representation, and the other based on physics-based site feature calculators.

- Physics-based site feature calculators: These strategies represent crystal structures by

calculating local geometric and chemical environment features. For instance, AGNIFingerprints calculate interatomic distances using Gaussian-weighted integrals, excelling in describing spatial arrangements and geometric shapes; OPSiteFingerprint focuses on describing coordination geometry and symmetry features of atoms; CrystalNNFingerprint integrates structural ordinal parameters and partial chemical information to enhance the representation of geometry and electronic interactions. However, these methods are limited in effectively capturing the symmetry, electronic and long-range interaction information of crystal structures.

- Data-driven structure graph representation strategies: These strategies leverage label feedback learning from large-scale training datasets to represent the geometric, symmetry, and long-range interaction information of crystal structures. For example, M3GNet and the CHGNet series perform better in multi-dimensional representation of crystal structures. Specifically, they use a VAE to compress KS wave functions(computed using ASE and mattersim calculators)[10, 27] into a low-dimensional latent space using an encoder and then reconstruct them using a decoder, efficiently representing electronic structures. By combining this with the design of multi-body interaction modules and label feedback from the training dataset, these models achieve a comprehensive representation of crystal structures. However, these strategies rely on the careful design of the VAE, GNN models, and datasets. For instance, MEGNet is less advanced in its model design compared to M3GNet and CHGNet, leading to suboptimal graph representation performance.

Figure 1b depicts our designed MLPSets framework, which aims to minimize the interference from GNN message-passing and graph convolution operations, thereby simplifying the GNN network design. This allows for a clear comparison of different crystal structure graph representation strategies to identify the most effective approach. The MLPSets framework takes various crystal structure graph representations as input, aggregates them into graph-level vectors, and passes them through multiple layers of MLPs. After processing by the MLPs, a readout function is applied to generate structure-level vectors, which are used for final material property prediction. This framework enables a more accurate evaluation of the performance of different representation strategies, facilitating the identification of the optimal representation method.

| Model name | Category | Description |
|---|---|---|
| MLPSets | reference | The standard node-edge constructed structural graph is used as input, passing through node or edge pooling layers and MLP layers to obtain predicted material properties as output. The Streamlined MLPSets framework, aiming to minimize the interference of GNN operations such as message passing and graph convolution, thereby simplifying the network design to clearly compare different crystal structure graph representation strategies and identify which strategy performs better. |
| AGNIFingerprints | Physics-based strategy | The structural graph, derived from crystal structure features calculated using physics-based AGNIFingerprints methods, is utilized as input. AGNIFingerprints describe the local environment of a site by computing the integral of the product of the radial distribution function and a Gaussian window function, taking into account Gaussian-weighted atomic distances and directional information. |
| OPSiteFingerprint | Physics-based strategy | The structural graph, derived from crystal structure features calculated using physics-based OPSiteFingerprint methods, is utilized as input. OPSiteFingerprint focuses on computing local structural order parameters for specific target structures (such as tetrahedra, octahedra, bcc). It evaluates the local geometry environment of a site by identifying the neighborhood shell that matches the expected coordination number. |
| CrystalNNFingerprint | Physics-based strategy | The structural graph, derived from crystal structure features calculated using physics-based CrystalNNFingerprint methods, is utilized as input. CrystalNNFingerprint combines structural order parameters (such as tetrahedrality, octahedrality) and chemical information (such as electron affinity differences) to describe the local environment of a site. |
| MEGNet | Data-driven strategy | Based on the MPtrj dataset, the structural graph, derived from crystal structure features learned through the pretrained MEGNET energy and force model, is utilized as input. The network design of MEGNet may not fully capture the complex interactions between atoms, or may not be as advanced in feature extraction as CHGNet and M3GNET. |
| M3GNET | Data-driven strategy | Based on the MPtrj dataset, the structural graph, derived from crystal structure features learned through the pretrained M3GNET energy and force model, is utilized as input. M3GNET can effectively represent geometric, symmetry, and long-range interaction information in crystal structures, but it lacks the capability to describe electronic structure information between sites within the structure. |
| CHGNet-NO-VAE | Data-driven strategy | Based on the MPtrj dataset, the structural graph, derived from crystal structure features learned through the pre-trained CHGNet energy and force model, is utilized as input. CHGNet does not incorporate electronic structure information generated by a VAE model based on Kohn-Sham wave functions. |
| CHGNet-V1 | Data-driven strategy | Based on the MPtrj dataset, the structural graph, derived from crystal structure features learned through the pre-trained CHGNet model for energy, force, stress, and magnetic moments, is utilized as input. By incorporating electronic structure information generated by a VAE model based on Kohn-Sham wave functions, CHGNet can provide rich information about local ion chemical environments and charge density distributions, as well as many-body effects, thereby enhancing the accuracy of material property predictions. |
| CHGNet-V2 | Data-driven strategy | Based on the DPA-2 training dataset, which includes 18 datasets covering metal alloys, battery materials, drug molecules, and ferroelectric materials across 73 elements, the structural graph is derived from crystal structure features learned through the pre-trained multi-task CHGNet model. By incorporating electronic structure information generated by a VAE model based on Kohn-Sham wave functions and leveraging multi-task material property labels, CHGNet can provide a comprehensive description of various dimensions of the crystal structure, especially the electronic structure and many-body effects, thereby enhancing the accuracy of material property predictions. |
| coGN | reference | The current state-of-the-art graph network models, coGN, are trained directly using each dataset. |
| MEGNet(kgcnn) | reference | A common reference, graph network models ,MEGNet, are trained directly on each dataset. |

Figure 2 presents the functional descriptions and classifications of two types of crystal structure graph representation strategies and reference models. The physics-based graph representation strategies includes AGNIFingerprints, OPSiteFingerprint, and CrystalNNFingerprint. The data-driven graph representation strategies encompasses MEGNet, M3GNET, CHGNet-NO-VAE, CHGNet-V1, and CHGNet-V2, with the best-performing strategy marked in green. The reference models include simple MLPSets and advanced GNN models such as coGN, where the crystal structure graph representations are based on a basic binary system of nodes and edges (i.e., elemental sites and the bonds between them).

The first category of crystal graph representation strategies is based on three physics-based site feature calculator models: AGNIFingerprints, OPSiteFingerprint, and CrystalNNFingerprint. AGNIFingerprints captures local geometric features by calculating Gaussian-weighted integrals of interatomic distances. This method excels in describing the spatial arrangement and local geometry of atoms, making it suitable for understanding structural stability and space-filling properties. However, it primarily focuses on geometric information and lacks electronic interactions. OPSiteFingerprint describes the local environment of atomic sites using local structural ordinal parameters such as tetrahedral and octahedral coordination. It effectively captures the coordination environment and local symmetry. However, it still falls short in describing electronic interactions. CrystalNNFingerprint integrates both geometric and chemical information, considering structural ordinal parameters and chemical information such as electron affinity differences and electronegativity. This approach provides a more comprehensive description of structure crystal, making it suitable for predicting properties involving complex chemical interactions. However, it may still require further optimization for handling very complex many-body and long-range interactions. Overall, CrystalNNFingerprint has a clear advantage in integrating geometric and chemical interactions, while AGNIFingerprints and OPSiteFingerprint excel in geometric structure description and structural ordinal parameter capture, respectively, but are limited in describing electronic interactions in crystal structures.

The second category of crystal graph representation strategies is based on five data-driven pre-trained GNN models: MEGNet, M3GNET, CHGNet-NO-VAE, CHGNet-V1, and CHGNet-V2. These models were pre-trained on the Materials Project Trajectory Dataset, which includes approximately 1.58 million atomic configurations and dynamic evolution information for 145,000 inorganic materials. MEGNet constructs the graph with atoms as nodes and chemical bonds as edges, updating features through graph convolution layers to capture local atomic interactions. However, it has several limitations: it provides a basic representation of geometric topological information, relying primarily on simple definitions of interatomic distances and bonds, which may not fully capture complex geometric features like bond bending, torsion, or non-covalent interactions. It lacks a multi-body interaction module, leading to a simplistic description of electronic interactions, mainly learned through feedback from material property labels during training. Additionally, it is inefficient in capturing long-range interactions, relying on increasing the number of graph convolution layers to extend the perception range. While MEGNet naturally handles permutation symmetry of atoms, it does not explicitly model symmetry operations (e.g., rotations, translations) as effectively as M3GNET and CHGNet.

M3GNET introduces a three-body interaction module to calculate atomic angles and dihedral angles, providing precise geometric topological descriptions. It learns electronic interaction indirectly through material property labels during training on the Materials Project Trajectory Dataset. M3GNET uses rotation-invariant feature representations and symmetry-preserving operations to ensure consistency of structural features under symmetry operations. It combines global state information (e.g., temperature, and charge) with node and edge features, achieving multi-scale feature learning from local atomic clusters to the entire crystal structure. However, its performance in predicting properties related to electronic structure and charge distribution is inferior to that of CHGNet, which includes magnetic moment information.

CHGNet Series (including CHGNet-NO-VAE, CHGNet-V1, and CHGNet-V2) precisely describes atomic angles and dihedral angles through a multi-body interaction module. It accurately characterizes electronic orbital occupation and atomic interactions by introducing constraints such as magnetic moments. The series ensures consistent structural features through rotation-invariant features and symmetry-preserving operations. It achieves multi-scale feature learning from local atomic environments to the global crystal structure through weighted message-passing mechanisms and multi-layer stacking, effectively capturing long-range interactions.

- CHGNet-NO-VAE maintains the same network architecture as CHGNet, pre-trained on the Materials Project Trajectory Dataset, but not incorporate electronic structure information generated by a VAE model based on KS wave functions in the crystal structure graph.
- CHGNet-V1 explicitly incorporates electronic structure information generated by a VAE model based on KS wave functions, as well as charge distribution information such as magnetic moments, as constraints for charge states. This allows for a more accurate description of the electronic orbital occupancy and electronic interactions of elements within the structure.
- CHGNet-V2 is based on the DPA-2 training dataset, which includes 18 datasets covering metal alloys, battery materials, drug molecules, and ferroelectric materials across 73 elements. It leverages indirect feedback from multi-task material property labels and incorporates electronic structure information generated by a VAE model based on KS wave functions. This approach enables a comprehensive representation of both geometric and electronic interactions in crystal structures, allowing for more accurate predictions of various material properties.

In summary, MEGNet captures local atomic interactions through graph convolution layers but has basic capabilities in representing geometric topology and long-range interactions. M3GNET introduces a three-body interaction module for precise geometric topological descriptions and uses rotation-invariant features and symmetry-preserving operations, combined with global state information for multi-scale crystal structure feature representation, but is less effective in handling electronic structure compared to CHGNet. The CHGNet series enhances the representation of electronic interactions through multi-body interaction modules and the introduction of an electronic structure generation model (VAE), along with a multi-objective label feedback strategy. This approach achieves a comprehensive representation of crystal structures, providing stronger generalization and adaptability.

## VAE for electronic structure representation

In this study, we employ a VAE model to generate electronic structure features based on KS wave functions, following the VAE method proposed by Bowen Hou[28]. They have carefully designed a dataset comprising 302 diverse samples based on the Computational 2D Materials Database (C2DB)[29-31], which includes various materials such as metals and semiconductors, with each material's unit cell containing 3 to 4 atoms. This dataset contains 68,384 DFT electronic states, which are randomly divided into a 90% training set and a 10% testing set. Using the KS wave function dataset and the VAE network architecture, the VAE model compresses the high-dimensional KS wave function magnitudes into a low-dimensional latent space using an encoder, and reconstructs the KS wave function magnitudes from the latent space using a decoder. This approach not only preserves the key physical information in the original data but also effectively represents the electronic structure.

We need to generate electronic structure representations for all structures in 13 benchmark datasets from Jarvis-DFT and Matbench using a VAE. If we were to use DFT calculations to generate the KS wave functions for each structure, it would be computationally expensive. Therefore, we employ the ASE and Microsoft's MatterSim-5M pre-trained interatomic potentials model to rapidly optimize the structures of the 13 benchmark datasets and generate electronic densities, similar to KS wave functions. Since the aim of this study is to compare the impact of different crystal structure graph representation strategies on model performance, highly precise KS wave function calculations are not necessary. It is important to emphasize that while the ASE and MatterSim-5M pre-trained interatomic potentials can generate electronic structure information similar to DFT calculations, they do not produce the KS wave functions themselves but rather the electronic densities and other related physical quantities. Notably, these models can be highly efficient alternatives to DFT calculations, especially when rapid predictions of physical properties for a large number of materials are required. However, for applications that require high-precision electronic structure information, DFT calculations remain indispensable.

# Model accuracies

| Property | MLPSets | MLPSets-AGNIFingerprints | MLPSets-OPSiteFingerprint | MLPSets-CrystalNNFingerprint | MLPSets-MEGNet | MLPSets-M3GNET | MLPSets-CHGNet-NO-VAE | coGN | MEGNet (kgcnn) | Data size |
|---|---|---|---|---|---|---|---|---|---|---|
| Trainable parameters | 60800 | 63872 | 65536 | 68608 | 64896 | 68992 | 77184 | 697601 | 155073 | * |
| Exfoliation energy | 42.1694 ± 10.77 | **36.938 ± 10.350** | 43.1846 ± 7.5648 | 40.3185 ± 8.765 | 42.0965 ± 8.102 | 39.0661 ± 7.898 | 39.1028 ± 7.889 | 37.1652 | 54.1719 | 636 |
| PhonDOS peak | 30.9347± 0.992 | 31.8315± 3.8021 | 27.2145 ± 2.0745 | 27.1047± 2.7311 | 28.8040 ± 1.579 | 28.7838 ± 2.852 | 30.2689± 2.019 | 29.7117 | 28.7606 | 1265 |
| dft_3d_dfpt_piezo_max_dielectric | 28.896±2.4797 | 29.0316±2.4388 | 28.5646 ± 2.4739 | 28.3673± 2.5626 | 28.5119±2.4294 | **28.3010±2.4253** | 28.3950± 2.1293 | 30.2923 | 30.1911 | 4704 |
| n | 0.3442 ± 0.0549 | 0.3343 ± 0.0550 | 0.3467±0.0550 | 0.3344 ± 0.0572 | 0.3252±0.0492 | 0.3251±0.0528 | **0.3211± 0.0501** | 0.3088 | 0.3391 | 4764 |
| dft_3d_exfoliation_energy | 47.6121±3.258 | **40.646±2.485** | 43.2510±5.5920 | 42.1808±4.7827 | 43.7876±5.1501 | 42.5065±4.825 | 42.1761±4.2454 | 47.6979 | 68.2435 | 812 |
| log(K_VRH) | 0.0709 ± 0.0015 | 0.0618 ± 0.0015 | 0.0631 ± 0.0022 | **0.0599 ± 0.0023** | 0.0654 ± 0.0007 | 0.0614 ± 0.0021 | 0.0610 ± 0.0011 | 0.0535 | 0.0668 | 10,987 |
| log(G_VRH) | 0.0970 ± 0.0018 | 0.0894 ± 0.0014 | 0.0883 ± 0.0024 | 0.0849 ± 0.00222 | 0.0912 ± 0.0018 | 0.0845 ± 0.0020 | **0.0839 ± 0.0015** | 0.0689 | 0.0871 | 10,987 |
| Perovskite | 0.0725±0.0027 | 0.0725±0.0016 | 0.0717±0.0022 | 0.0644 ±0.0035 | 0.0615±0.0017 | 0.0575 ±0.0019 | **0.0541 ± 0.0011** | 0.0269 | 0.0352 | 18,928 |
| dft_3d_mbj_bandgap | 0.4354±0.0199 | 0.4382±0.0209 | 0.4180±0.0093 | 0.3999±0.0104 | 0.4013±0.0093 | 0.3621±0.0128 | **0.3608±0.0123** | 0.264 | 0.297 | 18167 |
| dft_3d_slme | 6.1506±0.1809 | 6.066±0.1304 | 6.1629±0.1786 | 5.8824±0.1622 | 5.7848±0.0842 | **5.5302±0.1126** | 5.5987±0.1013 | 4.4507 | 5.0614 | 9062 |
| dft_3d_spillage | 0.3789±0.0120 | 0.3811±0.0126 | 0.3844±0.0116 | 0.3725±0.0110 | 0.3752±0.0091 | 0.3711±0.0087 | **0.3621± 0.0065** | 0.3609 | 0.3904 | 11375 |
| dft_3d_n_powerfact | 554.0408±3.535 | 550.5393±2.6031 | 556.2805±9.9119 | 540.1486±7.5776 | 533.3555±8.915 | 534.9692±6.023 | **530.4904±6.051** | 452.235 | 501.3722 | 23210 |
| dft_3d_magmom_oszicar | 0.4018±0.0100 | 0.3928±0.0099 | 0.3822±0.0064 | 0.3730± 0.0051 | 0.3898±0.0064 | 0.3451±0.0075 | **0.3391±0.0049** | 0.2502 | 0.3543 | 52210 |

Figure 3 presents a comparative analysis of two crystal structure graph representation strategies—physics-informed and data-driven—based on the MLPSets framework, along with reference models (such as coGN and MEGNet) on 13 benchmark datasets. All models were evaluated using a consistent five-fold random splitting method. For regression tasks, the figure shows the mean and standard deviation of the MAE, organized by dataset size. The best-performing models within the standard deviation for each dataset are highlighted in bold, while the second-best results are lightly shaded. The datasets used are obtained from Jarvis-DFT and Matbench.

| Property | MLPSets-CHGNet-V1 | MLPSets-CHGNet-V2 | coGN | MEGNet (kgcnn) | Data size |
|---|---|---|---|---|---|
| Trainable parameters | 77184 | 77184 | 697601 | 155073 | * |
| Exfoliation energy | 40.1028 ± 8.4389 | **36.092± 8.4389** | 37.1652 | 54.1719 | 636 |
| PhonDOS peak | 32.2689± 2.3439 | **30.3327± 2.3439** | 29.7117 | 28.7606 | 1265 |
| dft_3d_dfpt_piezo_max_dielectric | 28.5670± 2.4293 | **26.85298± 2.4293** | 30.2923 | 30.1911 | 4704 |
| n | 0.3199± 0.0569 | **0.3079± 0.0552** | 0.3088 | 0.3391 | 4764 |
| dft_3d_exfoliation_energy | 40.9761±4.6485 | **38.878±4.6485** | 47.6979 | 68.2435 | 812 |
| log(K_VRH) | 0.0610 ± 0.0011 | **0.0549± 0.0011** | 0.0535 | 0.0668 | 10,987 |
| log(G_VRH) | 0.0824 ± 0.0012 | **0.0742± 0.0012** | 0.0689 | 0.0871 | 10,987 |
| Perovskite | 0.0441 ± 0.0009 | **0.0391± 0.0009** | 0.0269 | 0.0352 | 18,928 |
| dft_3d_mbj_bandgap | 0.3187±0.0101 | **0.2868±0.0101** | 0.264 | 0.297 | 18167 |
| dft_3d_slme | 5.4124±0.1044 | **4.8720±0.1044** | 4.4507 | 5.0614 | 9062 |
| dft_3d_spillage | 0.3730± 0.0072 | **0.3359± 0.0072** | 0.3609 | 0.3904 | 11375 |
| dft_3d_n_powerfact | 525.4904±6.6605 | **472.974±6.6605** | 452.235 | 501.3722 | 23210 |
| dft_3d_magmom_oszicar | 0.3317±0.0053 | **0.2985±0.0053** | 0.2502 | 0.3543 | 52210 |

Figure 4 presents a comparative analysis of the best data-driven crystal structure graph representation strategies, CHGNet-V1 and CHGNet-V2, along with reference models on 13 benchmark datasets using the MLPSets framework. All models were evaluated using a consistent five-fold random splitting method. For regression tasks, the figure shows the mean and standard deviation of the MAE, organized by dataset size. The best-performing models within the standard deviation for each dataset are highlighted in bold, while the second-best results are lightly shaded. The datasets used are obtained from Jarvis-DFT and Matbench.

The analysis of Figure 3 reveals that, compared to MLPSets, a network design without complex GNN components (such as equivariant modules and message-passing blocks) and based on a simple node-edge binary system (i.e., elemental sites and their bonding), both the physics-based and data-driven crystal structure graph representation strategies within the MLPSets framework perform better on most datasets with varying data sizes. Further analysis of Figure 3 shows that, for small datasets (with data sizes ranging from 636 to 4764), the advanced graph representation strategies based on the MLPSets framework exhibit significant performance advantages over current reference models such as coGN and MEGNet. Despite having significantly fewer training parameters than these reference models, the MLPSets-based models effectively leverage advanced physics-based and data-driven crystal structure graph representation strategies to achieve accurate predictions of material properties. For larger datasets (with data sizes ranging from 9062 to 52210), although advanced models like coGN lead on all larger datasets, MLPSets-M3GNET and MLPSets-CHGNet outperform MEGNet on specific datasets, including log(KVRH), log(GVRH), dft_3d_spillage, and dft_3d_magmom_oszicar. This indicates that, while MEGNet is a powerful model, MLPSets-M3GNET and MLPSets-CHGNet achieve lower computational costs and better performance on certain datasets through more comprehensive crystal structure graph representation strategies. Supplementary Figures S2 to S8 show the learning curves of the MLPSets series models on 13 benchmark datasets sampled from Matbench and Jarvis-DFT. These figures display the changes in loss, MAE, and root mean squared error (RMSE) during both the training and testing phases. All results are based on 5-fold cross-validation to ensure the accuracy and reliability of the model's evaluation.

We found that data-driven crystal structure representation strategies (MLPSets-MEGNet, MLPSets-M3GNET, and MLPSets-CHGNet) significantly outperform physics-based representation strategies (MLPSets-AGNIFingerprints, MLPSets-OPSiteFingerprint, and MLPSets-CrystalNNFingerprint). Although the data-driven strategies generally do not match the performance of the current state-of-the-art model coGN on most datasets, they come close to the performance of MEGNet (kgcnn), with MLPSets-CHGNet-NO-VAE showing the best performance. This is primarily attributed to the advanced model design of CHGNet. Through the comparative analysis in Figure 3, we isolated the effects of GNN network design (including complex equivariant, many-body interaction, and message-passing mechanisms) by using the MLPSets framework. This allowed us to determine the effectiveness of advanced graph representation strategies and identify the data-driven CHGNet as the optimal approach. Based on this, we carefully designed the CHGNet-V1 and CHGNet-V2 graph representation strategy. When tested on 13 benchmark datasets using the MLPSets framework, as shown in Figure 4, both CHGNet-V1 and CHGNet-V2 demonstrated significant improvements over MLPSets-CHGNet-NO-VAE across all data sizes. They significantly outperformed MEGNet and comprehensively surpassed the current state-of-the-art model coGN on small datasets, while closely matching coGN's performance on large datasets. These enhancements are attributed to the careful design of the data-driven GNN models, the training datasets, and the introduction of the VAE model for generating electronic structure information.

The results from Figures 3 and 4 emphasize that achieving a comprehensive representation of crystal structures can significantly enhance model performance, even in the context of the MLPSets framework, which does not incorporate any complex equivariant, many-body interaction, or message-passing mechanisms. This comprehensive representation not only leads to substantial improvements in model performance but also significantly reduces the number of training and inference parameters, thereby further enhancing the model's generalization and robustness. Furthermore, the comparative tests between physics-based and data-driven crystal structure graph strategies demonstrate that data-driven approaches have a clear advantage in comprehensively describing crystal structures. These data-driven strategies are also the most promising methods for addressing the challenges of representing electronic interactions.

**Model convergence**

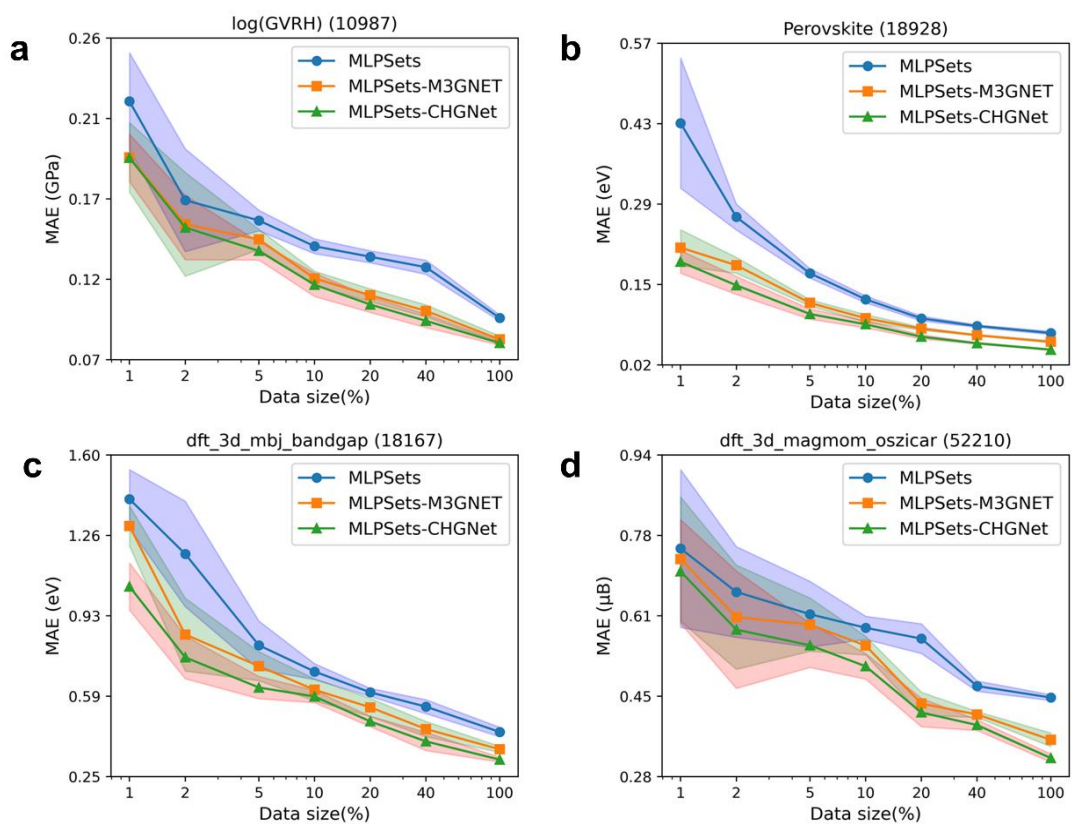

Figure 5 illustrates the convergence of the models MLPSets, MLPSets-M3GNET, MLPSets-CHGNet (specifically MLPSets-CHGNet-V1, referred to as MLPSets-CHGNet), and the reference model coGN. Panels (a) and (b) show the results for the log(GVRH) and Perovskite datasets from Matbench, while panels (c) and (d) correspond to the dft_3d_mbj_bandgap and dft_3d_magmom_oszicar datasets from Jarvis-DFT. The x-axis is plotted on a logarithmic scale to enhance resolution at smaller data sizes. The shaded regions represent the standard deviation across five random data fits.

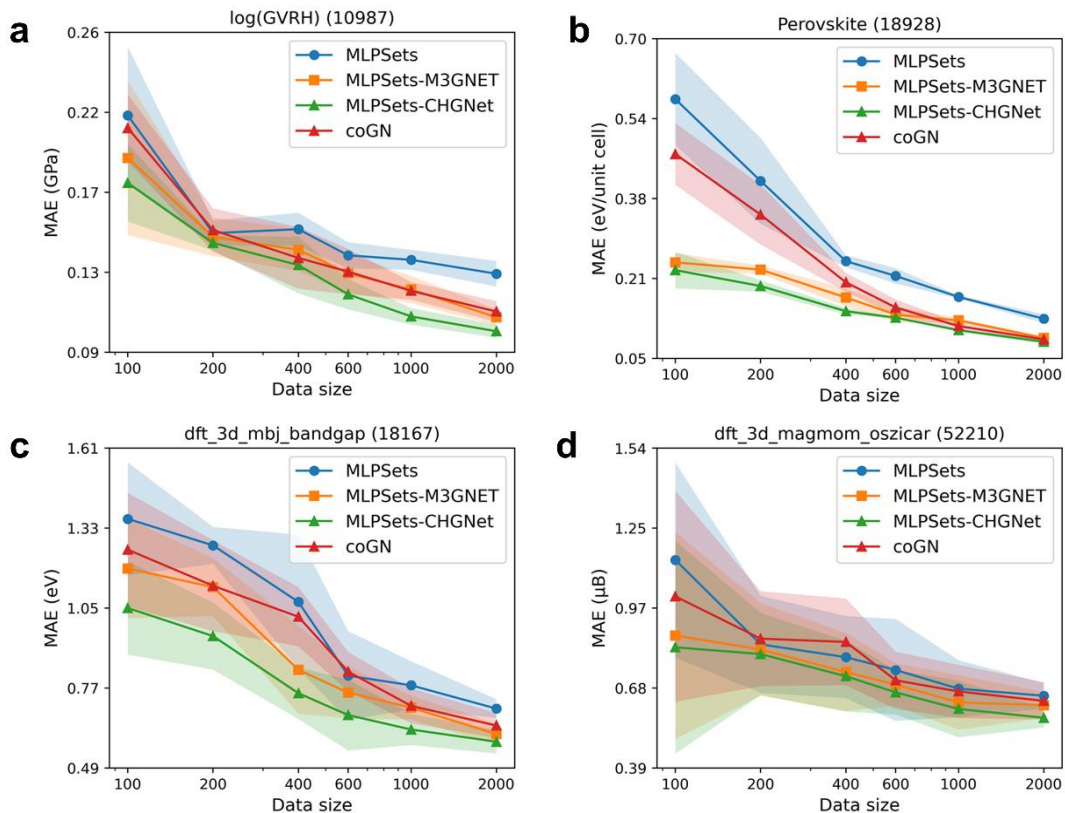

Figure 6 presents a study of the model convergence of MLPSets, MLPSets-M3GNET, MLPSets-CHGNet, and the reference model coGN under the small data limit. The four datasets include: (a) the log10 value of the shear modulus, (b) the calculated perovskite formation energy, (c) the bandgap with the Tran-Blaha modified Becke-Johnson (MBJ) potential, and (d) the magnetic moment dataset from Matbench and Jarvis-DFT. The shaded regions represent the standard deviation across five random data fits.

To evaluate the convergence performance of the models, we conducted a convergence study comparing MLPSets, MLPSets-M3GNET, MLPSets-CHGNet, and the advanced reference model coGN. As shown in Figure 5, these models were tested on the log(GVRH) and Perovskite datasets from Matbench, as well as the dft_3d_mbj_bandgap and dft_3d_magmom_oszicar datasets from Jarvis-DFT. The results indicate that the MLPSets-CHGNet model not only exhibits higher performance across all tasks but also converges faster compared to the reference model MLPSets. Furthermore, we focused on the model performance under limited data conditions, as shown in Figure 6. We selected the log(GVRH), Perovskite, dft_3d_mbj_bandgap, and dft_3d_magmom_oszicar datasets for testing. In this extreme test, MLPSets-CHGNet consistently outperformed the reference models coGN and MLPSets across different dataset sizes. Moreover, the predictive accuracy and stability of the model improved with an increase in the available data. For datasets with fewer than 400 samples, MLPSets-CHGNet exhibited significantly lower MAE on the Perovskite and dft_3d_mbj_bandgap datasets compared to the reference models coGN and MLPSets. This result further confirms the strong and efficient learning capability of MLPSets-CHGNet. The superior performance and rapid convergence of MLPSets-CHGNet on small datasets validate the effectiveness and stability of the data-driven crystal structure graph representation strategy based on CHGNet-V1. This performance can be attributed to the innovative GNN architecture and many-body interaction modules in CHGNet, which comprehensively represent key information in crystal structures. Additionally, the incorporation of magnetic moment information as a charge state constraint and the integration of electronic structure information generated by a VAE model based on KS wave functions enable accurate representation of electronic interactions. These features allow the model to simulate complex atomic interactions, thereby enhancing its overall performance.

## Model extrapolability

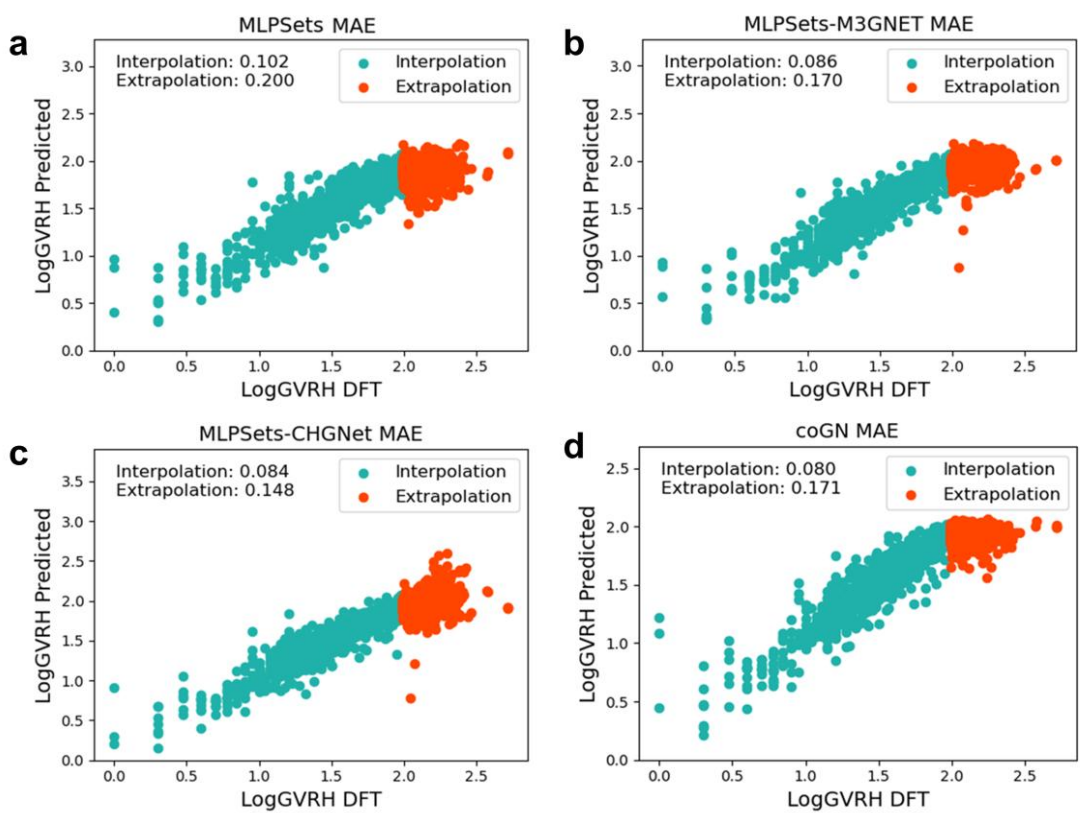

Figure 7 compares the absolute differences between the interpolation and extrapolation predictions of log(GVRH) and the DFT-calculated values. Panels (a) to (d) show the results for MLPSets, MLPSets-M3GNET, MLPSets-CHGNet, and the reference model coGN, respectively. The training and validation data were randomly sampled from the 0-90% target quantile range. Half of the test data comes from the 90-100% quantile range (extrapolation), while the other half is from the same target range as the training-validation data (interpolation).

In typical materials design problems, the goal is not to find materials with properties similar to existing ones, but to discover new materials with unique properties that extend beyond the current material library. This extrapolation task poses a significant challenge for most machine learning models. Previous studies often use leave-one-cluster-out cross-validation (LOCO CV)[32] or k-fold forward cross-validation[33] to evaluate model performance in extrapolation tasks, where validation is performed on data outside the training range. Here, we adopt the concept of forward cross-validation, dividing the data based on the target value range, and apply this method to the Perovskites, JarvisMbjBandgap, JarvisMagmomOszicar, log(GVRH), and log(KVRH) datasets. For example, using elasticity data (log(KVRH) and log(GVRH)), we can simulate the process of finding super-incompressible (high K) and super-hard (high G) materials. The top 10% of materials with the highest target values are used as the test dataset (high test set, extrapolation).

The remaining data is then divided into training, validation, and test sets (low test set, interpolation).

Figure 7 compares the prediction performance of MLPSets, MLPSets-M3GNET, MLPSets-CHGNet, and the advanced reference model coGN on the DFT-calculated shear modulus (log(GVRH)) dataset, showcasing their extrapolation capabilities in both interpolation and extrapolation tasks. The results show that the MLPSets-CHGNet model significantly outperforms the other models, including the complex coGN model, especially in extrapolation tasks. Notably, despite its simpler architecture and fewer parameters, MLPSets-CHGNet demonstrates superior extrapolation performance compared to the more parameter-intensive and structurally complex coGN model. This further validates the effectiveness of data-driven CHGNet-V1 graph representation strategies in enhancing model extrapolation performance.

Additionally, supplementary Figures S9 to S12 provide further confirmation of MLPSets-CHGNet's robust and stable extrapolation capabilities across different material property prediction tasks, including Perovskite, JarvisMbjBandgap, JarvisMagmomOszicar, and log(KVRH) datasets. When the prediction range is farther from the training data range, the accuracy of the MLPSets-CHGNet model, which uses magnetic moment information as charge state constraints, improves more significantly. These findings highlight the potential of MLPSets-CHGNet in discovering novel materials. This also emphasize the significance of incorporating magnetic moment data as constraints on charge states, as well as integrating electronic structure information produced by a VAE model to enhance the representation of structural electronic interactions. This data-driven crystal structure graph representation strategy is essential for enhancing the model's extrapolation capabilities.

# Key point identification in crystal structure graph representation

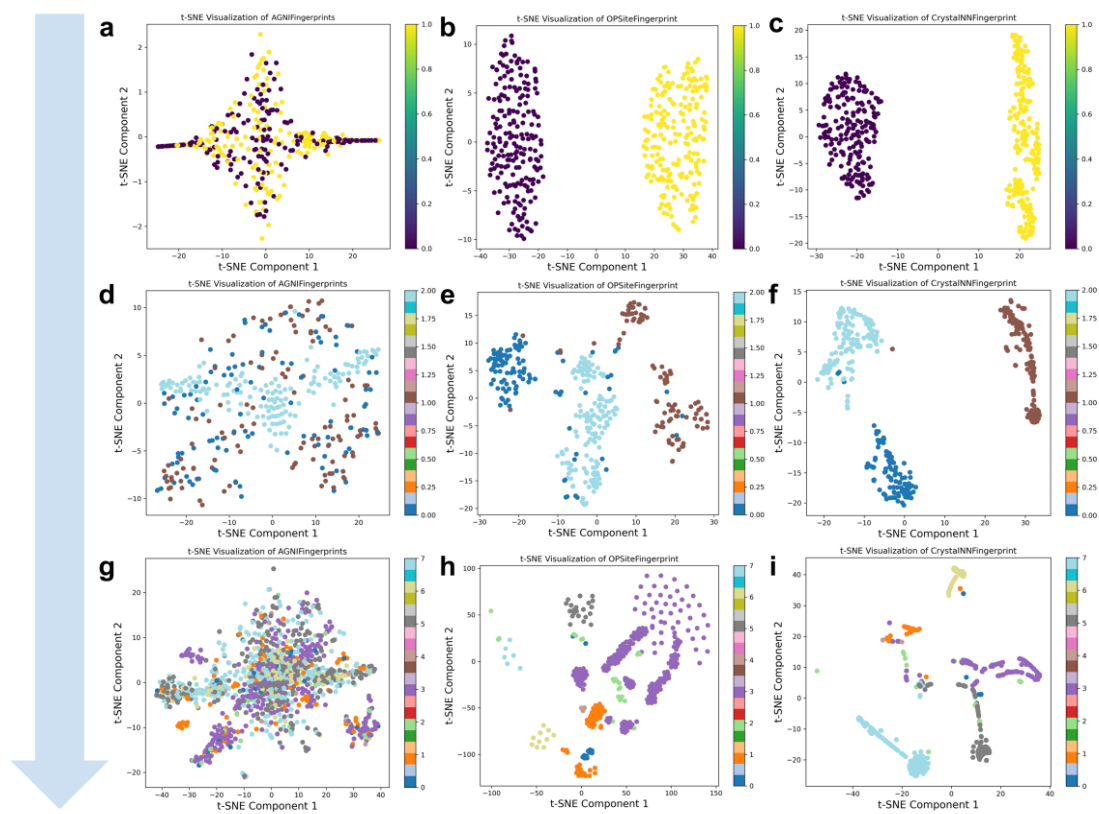

Figure 8, Multiscale classification of crystal structures based on physics-based features.
a-c: Present the t-SNE plots of the feature vectors for liquid and amorphous structures calculated by AGNIFingerprints, OPSiteFingerprint, and CrystalNNFingerprint, respectively.
d-f: Classification of carbon allotropes of different dimensions: 0D (clusters), 2D (sheets), and 3D (bulk). t-SNE plots of feature representations for the test dataset calculated by AGNIFingerprints, OPSiteFingerprint, and CrystalNNFingerprint, respectively. g-i: Classification of crystal structures based on space groups. t-SNE plots of feature space embeddings calculated by AGNIFingerprints, OPSiteFingerprint, and CrystalNNFingerprint, respectively.

Figure 8 demonstrates the multi-scale classification of crystal structures using physics-based features across three different levels: liquid and amorphous structures[34, 35], carbon allotropes of different dimensions (0D aggregates, 2D sheets, 3D bulk)[36-39], and crystal structure classification based on space groups[40]. This classification method progressively refines the distinction of crystal structures from macro to micro and from global to local. In the figure, a-g, b-h, and c-i show the multi-scale classification results of AGNIFingerprints, OPSiteFingerprint, and CrystalNNFingerprint, respectively, at these three different levels. The comparison reveals the following:

1. AGNIFingerprints performs poorly across all three levels (liquid and amorphous structures, carbon allotropes of different dimensions, and space groups). This indicates that relying solely on Gaussian-weighted integrals of interatomic distances and directions may not be sufficient to capture the geometric characteristics of each site in the crystal structure, including the topological structure and symmetry of its neighbors.

2. OPSiteFingerprint and CrystalNNFingerprint both achieve relatively accurate

classification across the three levels. This demonstrates the effectiveness of structural order parameters in describing crystal structures.

Although the figure8 shows that OPSiteFingerprint and CrystalNNFingerprint perform comparably across the three levels, the crystal structure graph representation strategy based on CrystalNNFingerprint is the most effective. This is because CrystalNNFingerprint, in describing the local environment, not only considers structural order parameters but also incorporates information about electronic interactions (such as electron affinity differences). This comprehensive approach, which considers both geometric and electronic interactions, which is crucial for predicting various material properties. In contrast, AGNIFingerprints primarily focus on Gaussian-weighted integrals of interatomic distances and directions, while OPSiteFingerprint emphasizes the calculation of local structural order parameters. Although these methods have their advantages in describing local structures, they do not consider electronic interaction information as comprehensively as CrystalNNFingerprint, which may explain why they perform less well in benchmark tasks.

We further compared the classification capabilities of physics-based and data-driven crystal structure representation strategies across three levels: liquid and amorphous structures, carbon allotropes of different dimensions, and crystal structures based on space groups. As shown in Figures S13-S15, while AGNIFingerprints performed poorly across all three levels, the other physics-based models demonstrated good classification performance. The data-driven crystal structure representation strategies performed well in classifying macroscopic and simple structures but showed poorer performance in the fine classification of structures based on space groups. This discrepancy suggests that data-driven graph representation features not only capture geometric topological patterns but also integrate information about elemental chemistry, geometry, electronic interactions, symmetry, and complex interactions, providing a more comprehensive representation of crystal structures. Therefore, although data-driven strategies may be less effective in distinguishing fine local geometric structures, they perform better in benchmark tests and practical applications. This indicates that a more comprehensive crystal structure graph representation is crucial, and a complete description of chemical information related to electronic interactions is more important than the fine differentiation of geometric structures at different scales. These findings are consistent with the research by Chi Chen et al[41]., who observed a significant increase in error across all datasets when using models with only structural inputs (without elemental information). Surprisingly, a classification model using only geometric information achieved an AUC value of 0.92 on the MP metallic properties dataset, indicating that geometric information alone can distinguish between metals and non-metals at a macroscopic level. However, the chemical information provided by elemental data, which includes electronic interactions, is more critical for predicting material properties.

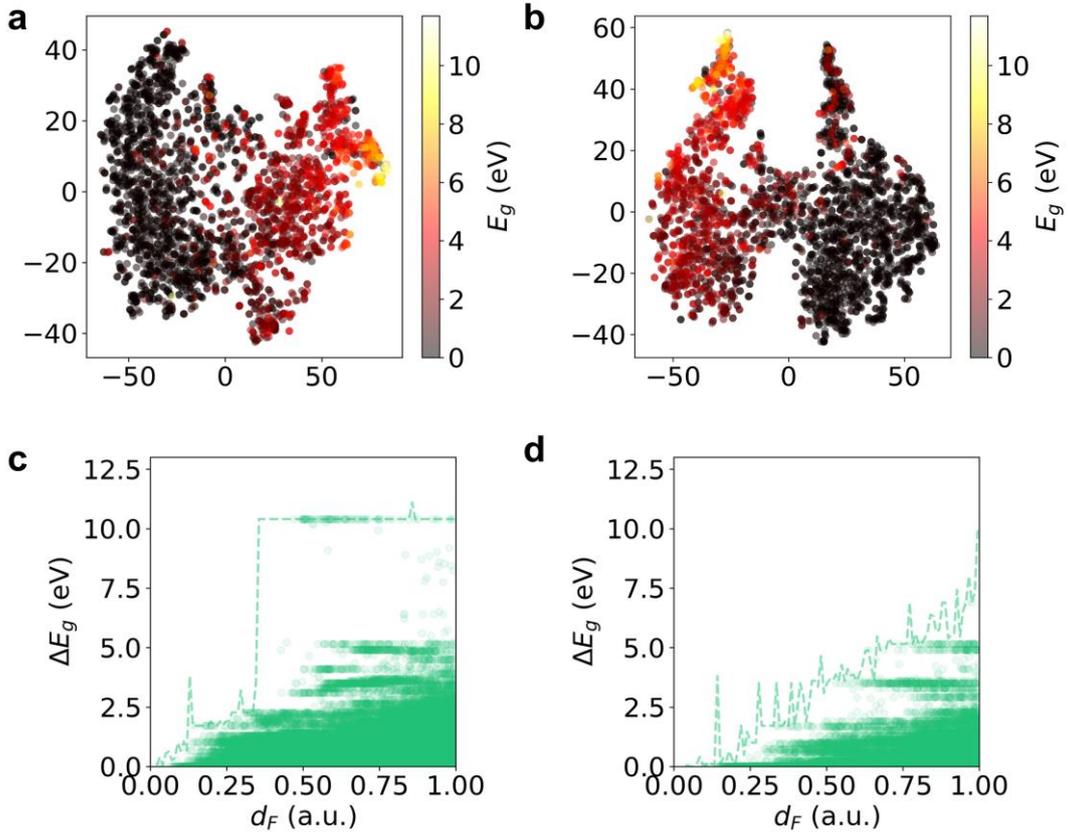

Figure 9, Comparison of MLPSets-M3GNET and MLPSets-CHGNet on latent structural features. a and b, Two-dimensional t-distributed stochastic neighbor embedding projection of features for MLPSets-M3GNET (a) and MLPSets-CHGNet (b) models trained using 2643 experimental data points. The markers are colored according to the experimental band gap. c and d, Plots of the experimental band gap difference ($\Delta E_g$) against normalized latent structural feature distance (dF) in arbitrary units (a.u.) for the MLPSets-M3GNET (c) and MLPSets-CHGNet (d) models trained on all available experimental data. The dashed lines indicate the envelope of the maximum $\Delta E_g$ at each dF. The scattering points are sub-sampled by a factor of 15.

Figure 9 reveals why the crystal structure representation strategy based on CHGNet outperforms M3GNET. Specifically, Figure 9a and 9b show the two-dimensional t-SNE[42] projections of the features for the MLPSets-M3GNET and MLPSets-CHGNet models, respectively, trained using 2643 experimental data points[43]. The markers are colored according to the experimental band gap. In Figures 9c and 9d, we plot the relationship between the experimental band gap difference ($\Delta E_g$) and the normalized latent structural feature distance (dF) for the MLPSets-M3GNET and MLPSets-CHGNet models, respectively. The dashed lines indicate the envelope of the maximum $\Delta E_g$ at each dF. The comparison shows that CHGNet produces superior structural representations, allowing for a clearer distinction between structures with large band gap differences. The CHGNet model exhibits an almost linear correspondence between dF and max $\Delta E_g$, especially when dF is less than 1. This indicates that CHGNet has higher resolution and accuracy in learning and representing structural features. In contrast, the M3GNET model has poorer resolution for experimental band gaps, particularly for large $\Delta E_g$ values. The range of dF from 0.25 to 1 corresponds to a maximum $\Delta E_g$ of about 10 electron volts, and the relationship between dF and max $\Delta E_g$ is very noisy at low values, indicating that

M3GNET struggles to distinguish structures with similar structural features but large band gap differences. We believe that CHGNet achieves better results primarily because it provides rich information about the local ionic environment and charge distribution by incorporating magnetic moment as a constraint on charge states and integrating electronic structure information generated by a VAE model based on KS wave functions. The inclusion of this information helps the model learn more effective latent structural features, thereby significantly improving the predictive performance of experimental high-fidelity bandgaps.

| Property | MLPSets-CHGNet | MLP-CHGNet-OPSiteFingerprint | %Improve | coGN | MEGNet Kgcnn | Data size |
|---|---|---|---|---|---|---|
| Trainable parameters | 77184 | 87680 | * | 697601 | 155073 | * |
| Exfoliation energy | 40.1028 ± 8.4389 | 38.099 ±7.4534 | 5.00 | 37.1652 | 54.1719 | 636 |
| PhonDOS peak | 32.2689 ± 2.3439 | 29.9049 ±2.553 | 7.33 | 29.7117 | 28.7606 | 1265 |
| dft_3d_dfpt_piezo_max_dielectric | 28.5670 ± 2.4293 | 28.4181 ±2.5455 | 0.52 | 30.2923 | 30.1911 | 4704 |
| n | 0.3199 ± 0.0569 | 0.2930 ±0.0513 | 8.41 | 0.3088 | 0.3391 | 4764 |
| dft_3d_exfoliation_energy | 40.9761 ±4.6485 | 40.6541 ±6.5071 | 0.79 | 47.6979 | 68.2435 | 812 |
| log($K_{VRH}$) | 0.0610 ± 0.0011 | 0.0551 ±0.0023 | 9.67 | 0.0535 | 0.0668 | 10,987 |
| log($G_{VRH}$) | 0.0824 ± 0.0012 | 0.0773 ±0.0018 | 6.19 | 0.0689 | 0.0871 | 10,987 |
| Perovskite | 0.0441 ± 0.0009 | 0.0434 ±0.0008 | 1.59 | 0.0269 | 0.0352 | 18,928 |
| dft_3d_mbj_bandgap | 0.3187 ±0.0101 | 0.3093 ±0.0096 | 2.95 | 0.264 | 0.297 | 18167 |
| dft_3d_slme | 5.4124 ±0.1044 | 5.1979 ±0.1199 | 3.96 | 4.4507 | 5.0614 | 9062 |
| dft_3d_spillage | 0.3730 ± 0.0072 | 0.3651 ±0.0069 | 2.12 | 0.3609 | 0.3904 | 11375 |
| dft_3d_n_powerfact | 525.4904 ±6.6605 | 516.947 ±6.7801 | 1.63 | 452.235 | 501.3722 | 23210 |
| dft_3d_magmom_oszicar | 0.3317 ±0.0053 | 0.3195 ±0.0039 | 3.68 | 0.2502 | 0.3543 | 52210 |

Figure 10 demonstrates the performance enhancement of the MLPSets-CHGNet-OPSiteFingerprint model, which combines data-driven and physics-based features from CHGNet and OPSiteFingerprint, across multiple material property datasets. The figure provides a detailed comparison of the performance differences between this combined model and the MLPSets-CHGNet model, which uses only CHGNet features, and presents the performance improvement in percentage form. Additionally, the figure compares the performance of the MLPSets-CHGNet-OPSiteFingerprint model with that of the current advanced reference models, MEGNet (kgcnn) and coGN, while also indicating the number of training parameters for each model.

After a thorough analysis of Figures 8 and 9, we recognize that when describing the local environment of materials, it is necessary to consider both structural order parameters (such as tetrahedrality and octahedrality) and chemical information (such as differences in electron affinity). Therefore, we combined the data-driven CHGNet-V1 features, which excel in representing electronic interaction information, with the physics-based OPSiteFingerprint features, which are adept at distinguishing crystal geometric topologies, to construct the MLPSets-CHGNet-OPSiteFingerprint model. This model aims to achieve a comprehensive representation of both geometric and electronic interactions in crystal structures. As shown in Figure 10, MLPSets-CHGNet-OPSiteFingerprint achieved performance improvements across all property datasets, regardless of their size, with the maximum improvement reaching approximately 10%. Notably, MLPSets-CHGNet-OPSiteFingerprint, using only 87680 training parameters, outperformed MEGNet (kgcnn), which uses 155073 training parameters, on 8 out of 13 datasets. Even when compared to the advanced coGN model, which uses 697601 training parameters, MLPSets-CHGNet-OPSiteFingerprint outperformed coGN on 3 datasets and was very close to coGN's performance on 7 datasets. This demonstrates the powerful learning efficiency and simplicity of MLPSets-CHGNet-OPSiteFingerprint and validates our analysis that the ability to describe atomic chemical interactions and distinguish geometric structures at different scales is crucial for predicting material properties. This points to the direction for future small-scale, high-efficiency model design in various scenarios involving different levels of data scarcity in material design and optimization. Figure S16 provides the learning curves of MLPSets-CHGNet-OPSiteFingerprint on the 13 benchmark datasets shown in Figure 10. To further demonstrate the superior performance of MLPSets-CHGNet-OPSiteFingerprint on small experimental datasets, Table S1 lists its performance relative to current advanced reference GNNs on experimental datasets covering 184 to 2647 samples. Clearly, MLPSets-CHGNet-OPSiteFingerprint shows significant advantages across all experimental datasets.

## Applications

**DenseGNN performance improved using CHGNet-V1-based structure graph representation**

| Property | DenseGNN-CHGNet | ALIGNN | DN++ (kgcnn) | M3GNET | MODNet | coGN | coNGN | Data size |
|---|---|---|---|---|---|---|---|---|
| Jdft2d | **32.4189 ± 8.329** | 43.4240± 8.949 | 42.6637± 13.720 | 50.1719± 11.90 | 33.192± 7.343 | 37.1652± 13.683 | 36.170± 11.597 | 636 |
| Phonons | **24.7388± 3.085** | 29.5390± 2.115 | 38.9636± 1.9760 | 34.1606± 4.500 | 34.2751± 2.078 | 29.7117± 1.997 | 28.887± 3.284 | 1265 |
| Dielectric | 0.2778±0.055 | 0.3449± 0.0871 | 0.3277± 0.0829 | 0.3120± 0.063 | **0.2711± 0.0714** | 0.3088± 0.0859 | 0.3142± 0.074 | 4764 |
| Perovskites | 0.0274 ± 0.0006 | 0.0288± 0.0009 | 0.0342± 0.0005 | 0.0330± 0.0001 | 0.0908± 0.0028 | **0.0269± 0.0008** | 0.0290± 0.0011 | 18,928 |
| Log gvrh | 0.0669± 0.0007 | 0.0715± 0.0006 | 0.0796± 0.0022 | 0.0860± 0.002 | 0.0731± 0.0007 | 0.0689± 0.0009 | **0.0670± 0.0006** | 10,987 |
| Log kvrh | 0.0508±0.0030 | 0.0568± 0.0028 | 0.0590± 0.0022 | 0.0580± 0.003 | 0.0548± 0.0025 | 0.0535± 0.0028 | **0.0491± 0.0026** | 10,987 |
| Band gap | **0.1371±0.0030** | 0.1861± 0.0030 | 0.2352± 0.0034 | 0.1830± 0.0050 | 0.2199± 0.0059 | 0.1559± 0.0017 | 0.1697± 0.0035 | 106,113 |
| Formation enthapy | 0.0175±0.0002 | 0.0215± 0.0005 | 0.0218± 0.0004 | 0.01950± 0.0002 | 0.0448± 0.0039 | **0.0170± 0.0003** | 0.01780± 0.0004 | 132,752 |

Figure 11, Comparison of test MAE results on Matbench datasets. This figure compares the Test MAE of Matbench datasets between DenseGNN-CHGNet and reference models including ALIGNN, DN++, coGN, M3GNET, MODNet, coGN, and coNGN. It is worth noting that coNGN, DN++, M3GNET, and ALIGNN belong to the nested graph networks category and incorporate angle information. MODNet is specifically optimized for small datasets by integrating domain knowledge. The best results are highlighted in bold. The figure also provides the data size for each property.

| Property | DenseGNN-CHGNet | coGN | coNGN | Schnet (kgcnn) | MEGNet (kgcnn) | DN++ (kgcnn) | ALIGNN | Data size | %Improve |
|---|---|---|---|---|---|---|---|---|---|
| dft_3d_mepsz | **22.4046** | 24.1081 | 22.842 | 25.668 | 27.292 | 30.3644 | 23.7313 | 16809 | **1.91** |
| dft_3d_exfoliation_energy | **34.7661** | 47.6979 | 46.272 | 48.3027 | 68.2435 | 46.1517 | 52.7033 | 812 | **24.67** |
| dft_3d_shear_modulus_gv | **8.1808** | 8.6612 | 8.4881 | 10.2291 | 10.4359 | 26.0817 | 9.476 | 19680 | **3.62** |
| dft_3d_spillage | 0.3514 | 0.3609 | **0.3463** | 0.3622 | 0.3904 | 0.4137 | 0.351 | 11375 | **-1.47** |
| dft_3d_optb88vdw_total_energy | 0.0263 | **0.0262** | 0.0273 | 0.0374 | 0.0393 | 0.051 | 0.0367 | 55713 | **-0.38** |
| dft_3d_mepsx | **23.3792** | 24.2289 | 23.3801 | 26.0289 | 26.6785 | 31.9568 | 24.0458 | 16809 | **0.0038** |
| dft_3d_epsz | 18.1902 | 19.6192 | **17.8104** | 21.5016 | 22.6781 | 33.8379 | 19.5678 | 44490 | **-2.13** |
| dft_3d_dfpt_piezo_max_dij | 15.4503 | 15.2235 | **13.8868** | 18.6753 | 15.7584 | 13.9889 | 20.5705 | 3345 | **-11.26** |
| dft_3d_mepsy | **23.2373** | 24.1891 | 23.3299 | 25.5455 | 25.9523 | 31.0215 | 23.6482 | 16809 | **0.40** |
| dft_3d_kpoint_length_unit | **9.3278** | 9.5722 | 9.3459 | 10.1022 | 10.3826 | 11.8875 | 9.5146 | 55392 | **0.19** |
| dft_3d_n_powerfact | **425.8875** | 452.235 | 456.6118 | 495.4136 | 501.3722 | 568.836 | 442.299 | 23210 | **3.71** |
| dft_3d_ph_heat_capacity | 7.7637 | 6.1125 | 7.8127 | 17.7707 | **6.0443** | 23.3618 | 9.6064 | 12054 | **-28.45** |
| dft_3d_formation_energy_peratom | **0.0268** | 0.0271 | 0.0291 | 0.0365 | 0.0423 | 0.0528 | 0.0331 | 55713 | **-1.10** |
| dft_3d_epsx | 19.2045 | 20.0004 | **18.5738** | 22.0798 | 21.775 | 27.2511 | 20.3942 | 44490 | **-3.40** |
| dft_3d_optb88vdw_bandgap | **0.1074** | 0.1219 | 0.1267 | 0.698 | 0.146 | 0.2247 | 0.1423 | 55713 | **11.90** |
| dft_3d_max_efg | **18.6633** | 20.4417 | 19.5495 | 23.4912 | 23.0652 | 26.9552 | 19.1211 | 11865 | **2.40** |
| dft_3d_epsy | 19.1386 | 19.7796 | **18.5923** | 22.1843 | 22.4066 | 34.1277 | 19.9987 | 44490 | **-2.94** |
| dft_3d_encut | **120.803** | 133.8915 | 129.8266 | 253.3669 | 139.6071 | 164.315 | 133.7962 | 55386 | **6.95** |
| dft_3d_n_Seebeck | **38.7830** | 39.2692 | 40.0977 | 47.244 | 47.2813 | 54.2759 | 40.9214 | 23210 | **1.24** |
| dft_3d_ehull | **0.0442** | 0.0466 | 0.0485 | 0.1014 | 0.0569 | 0.3685 | 0.0763 | 55364 | **5.15** |
| dft_3d_bulk_modulus_kv | 8.8267 | 8.992 | **8.7022** | 10.7105 | 11.4287 | 13.3743 | 10.3988 | 19680 | **-1.43** |
| dft_3d_avg_hole_mass | **0.1176** | 0.1372 | 0.1285 | 0.1459 | 0.1416 | 0.1709 | 0.1239 | 17642 | **5.08** |
| dft_3d_avg_elec_mass | **0.0778** | 0.0917 | 0.0876 | 0.0866 | 0.0896 | 0.112 | 0.0853 | 17642 | **8.79** |
| dft_3d_mbj_bandgap | **0.2398** | 0.264 | 0.2719 | 0.3289 | 0.297 | 0.4764 | 0.3104 | 18167 | **9.17** |
| dft_3d_dfpt_piezo_max_dielectric | 27.8295 | 30.2923 | **25.5553** | 26.5276 | 30.1911 | 30.3358 | 28.1514 | 4704 | **-8.89** |
| dft_3d_slme | **4.2315** | 4.4507 | 4.4428 | 5.3222 | 5.0614 | 5.6403 | 4.5207 | 9062 | **4.76** |
| dft_3d_magmom_oszicar | **0.2370** | 0.2502 | 0.2437 | 0.7755 | 0.7753 | 0.3995 | 0.2574 | 52210 | **2.75** |

Figure 12, Comparison of test MAE results on JARVIS-DFT datasets. This figure presents the Test MAE comparison of JARVIS-DFT datasets between DenseGNN-CHGNet and reference models including coGN, coNGN, Schnet, MEGNet, DimNetPP (DN++), and ALIGNN. It is worth noting that coNGN, DN++, and ALIGNN belong to the nested graph networks category and incorporate angle information. The best results and relative improvements are highlighted in bold. The figure also provides the data size for each property.

In Figures 11 and 12, we demonstrate the superior performance of the DenseGNN-CHGNet model on the MatBench and JARVIS-DFT datasets, utilizing the crystal structure graph representation strategy CHGNet-V1 identified in this paper. CHGNet-V1 is based on magnetic moment information serving as a charge state constraint and incorporates electronic structure information generated by a VAE model based on KS wave functions. Through extensive testing across 35 different benchmark datasets, DenseGNN-CHGNet not only competes favorably with current state-of-the-art models but also achieves significant leading in many cases. To ensure fair and accurate assessment, meticulous measures were taken to maintain consistency in comparisons. All models were trained on the same datasets, using appropriate training methods, optimal hyperparameters, and a unified cross-validation approach. The implementations of these models were based on the Keras Graph Convolutional Neural Network (KGCNN) framework[44], with hyperparameters set via a standardized JSON configuration file, ensuring reproducibility of experiments. Supplementary Figure S17 provides an overview of the DenseGNN architecture. Figures S18 and S19 present the learning curves of the DenseGNN-CHGNet model on the MatBench and JARVIS-DFT datasets, displaying the changes in loss, MAE, and RMSE during both the training and testing phases. All results are based on 5-fold cross-validation to ensure the accuracy and reliability of the model's evaluation.

Figure 11 reveals the performance of DenseGNN-CHGNet on eight regression task datasets from MatBench. DenseGNN-CHGNet exhibits superior performance in most tasks, particularly on jdft2d, phonons, log_kvrh, and band gap datasets, showing notable advantages over models such as ALIGNN[14], DN++[45], SchNet[46], M3GNET[2], MODNet[47], coGN[23], and coNGN[23] from previous models. It is worth noting that DenseGNN-CHGNet does not employ complex nested graph network architectures or domain-specific knowledge to enhance its performance. Instead, it effectively captures the geometric topology and electronic interactions of crystal structures by integrating Dense Connected Networks (DCNs) and Hierarchical Node-Edge-Graph Residual Networks (HRNs), along with leveraging the data-driven structural graph representation strategy CHGNet-V1. Figure 12 showcases the performance of DenseGNN-CHGNet on the JARVIS-DFT dataset, where it outperforms the coNGN model, which utilizes intricate nested graph network architectures, in most material property prediction tasks. This further validates the effectiveness of the data-driven crystal structure graph representation strategy CHGNet-V1 in enhancing model performance.

Overall, these results not only demonstrate the potential of DenseGNN-CHGNet as a novel graph neural network for material property prediction but also highlight the importance of the data-driven crystal structure graph representation strategy, CHGNet-V1, in enhancing model generalization and predictive accuracy.

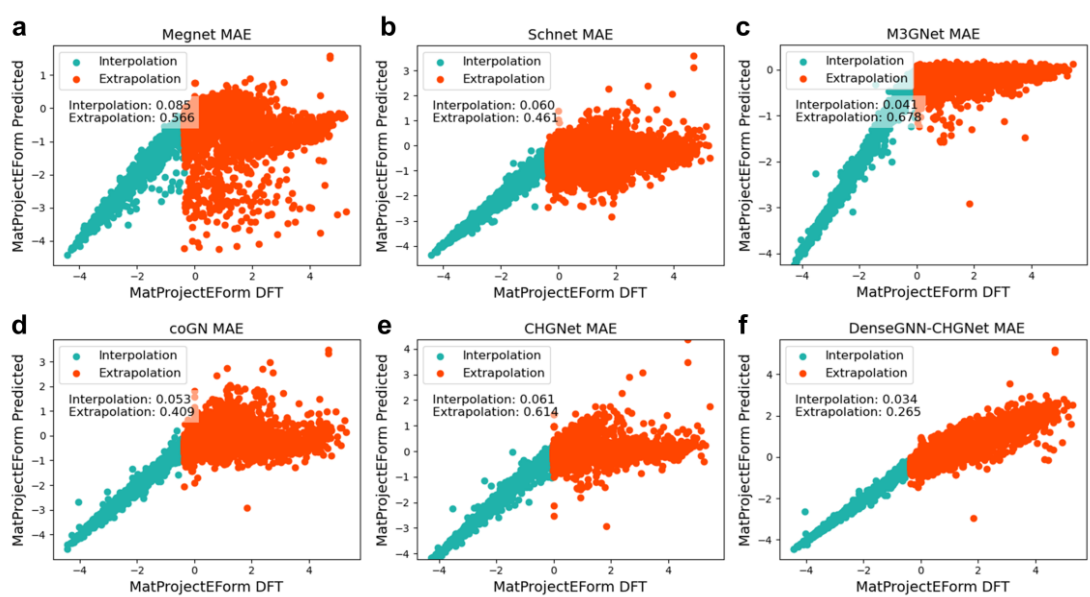

Figure 13 illustrates the absolute difference between the interpolated and extrapolated predicted values and the DFT-calculated values for the Material Project formation energy per atom. Figures 13a-13f present the extrapolation test results for MEGNet, Schnet, M3GNET, coGN, CHGNet, and DenseGNN-CHGNet, respectively. The training and validation data were randomly selected from the 0th to 70th percentile range of the target values. Half of the test data were drawn from the 70th to 100th percentile range (extrapolation), while the remaining half were from the same target range as the training-validation data (interpolation).

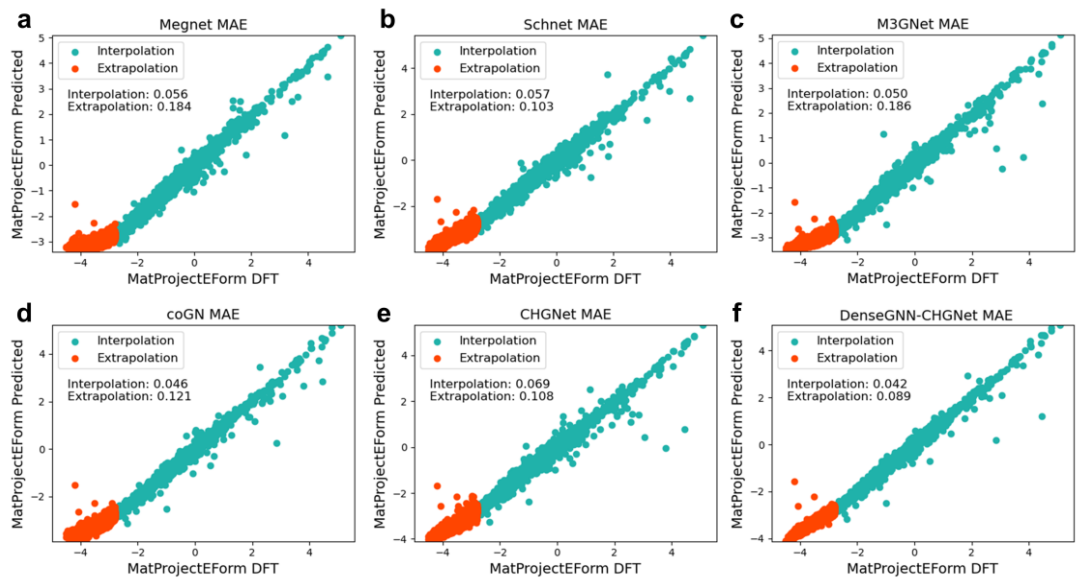

Figure 14 illustrates the absolute difference between the interpolated and extrapolated predicted values and the DFT-calculated values for the Material Project formation energy per atom. Figures 14a-14f present the extrapolation test results for MEGNet, Schnet, M3GNET, coGN, CHGNet, and DenseGNN-CHGNet, respectively. The training and validation data were randomly selected from the 30th to 100th percentile range of the target values. Half of the test data were drawn from the 0th to 30th percentile range (extrapolation), while the remaining half were from the same target range as the training-validation data (interpolation).

In typical material design problems, the goal is to discover novel materials with unique properties beyond the current material library, rather than simply finding materials with properties similar to existing ones. This extrapolation task poses a significant challenge for most machine learning models. To validate the effectiveness of the data-driven crystal structure graph representation strategy CHGNet-V1, we employed a forward cross-validation approach, dividing the data based on the target value range and applying it to the Material Project formation energy per atom dataset to test the extrapolation performance of DenseGNN-CHGNet.

Figure 13 illustrates the prediction performance of MEGNet, Schnet, M3GNET, coGN, CHGNet, and DenseGNN-CHGNet in both interpolation and extrapolation tasks. Training and validation data were randomly selected from the 0th to 70th percentile range of the target values. Half of the test data came from the 70th to 100th percentile range (extrapolation), while the other half came from the same target range as the training-validation data (interpolation). The results show that DenseGNN-CHGNet significantly outperforms other models in terms of prediction accuracy, particularly in extrapolation tasks, where its extrapolation performance surpasses that of the current advanced models coGN and CHGNet. This further validates the effectiveness of the data-driven crystal structure graph representation strategy CHGNet-V1 in enhancing the model's extrapolation capabilities. Figure 14 shows the prediction performance of the same six models in both interpolation and extrapolation tasks. Training and validation data were randomly selected from the 30th to 100th percentile range of the target values. Half of the test data came from the 0th to 30th percentile range (extrapolation), while the other half came from the same target range as the training-validation data (interpolation). The results show that DenseGNN-CHGNet also significantly outperforms other models in terms of prediction accuracy, particularly in extrapolation tasks, where its extrapolation performance again surpasses that of the current advanced models coGN and CHGNet. This further validates the effectiveness of the data-driven crystal structure graph representation strategy CHGNet-V1 in improving the model's extrapolation capabilities.

In summary, DenseGNN-CHGNet demonstrates excellent performance in two different extrapolation test settings, proving the effectiveness of the data-driven crystal structure graph representation strategy CHGNet-V1 in enhancing the model's extrapolation capabilities.

**DenseGNN based on CHGNet-V1's structure graph representation provides higher-fidelity pre-training data**

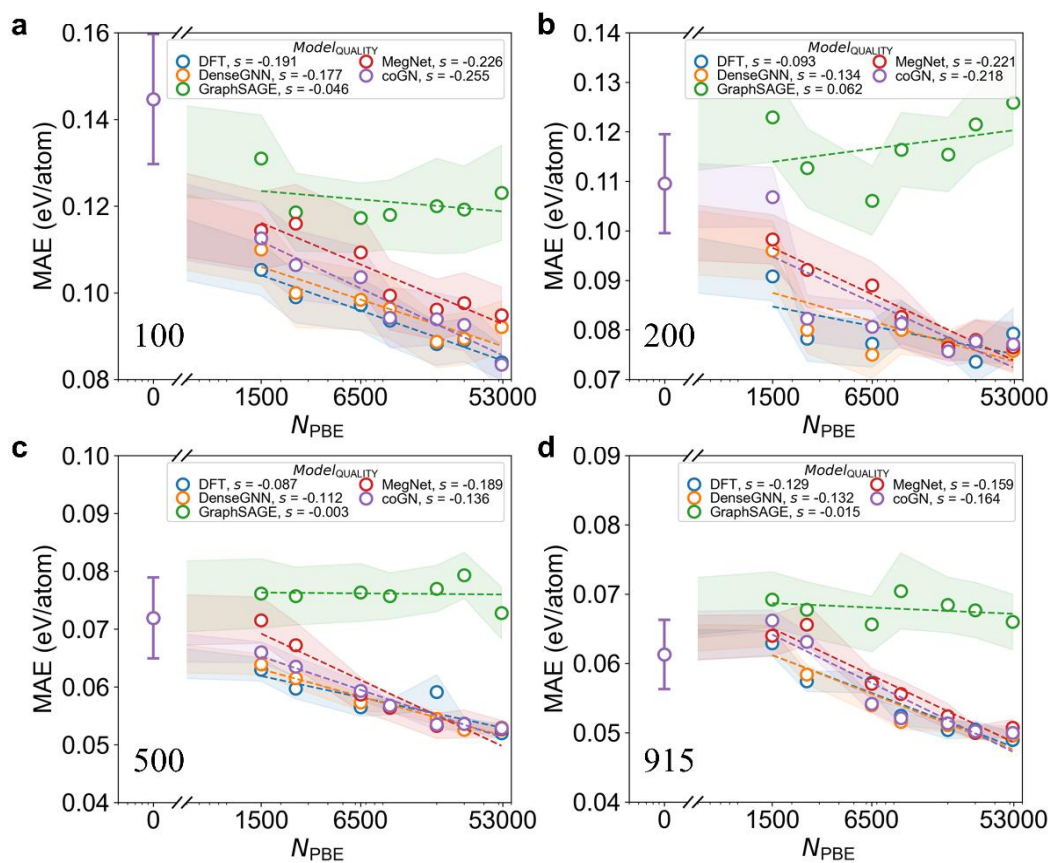

Figure 15, The error bars represent the standard deviation range, while points indicate model errors. Panels a-d showcase the performance of five models, namely GraphSAGE, MEGNet, coGN, DenseGNN-CHGNet, and DFT, which provided different fidelity pre-training formation energy data for MLPSets-CHGNet. These pre-trained models were then transferred to experimental formation energy datasets[43] containing 100, 200, 500, and 915 samples. As the fidelity of the pre-training data and data size increases, the error lines decrease, with the DenseGNN-CHGNet exhibiting the most accurate performance closest to DFT. The x-axis is plotted on a logarithmic scale, and the shaded regions represent one standard deviation of the MAE. The slope s indicates the linear fit of MAE with respect to $\log_{10} N_{PBE}$. Notably, GraphSAGE is an exception, as it provides or predicts data with very low fidelity, and its error lines do not decrease significantly despite the increase in data size.

In Figures 15a-d, we evaluated the impact of GraphSAGE[48], MEGNet, coGN, DenseGNN-CHGNet, and DFT-provided predictive data with varying fidelity and scale on the performance of the MLPSets-CHGNet model when transferred to small-scale experimental datasets. The fidelity of these data ranged from low to high, approaching the precision of DFT calculations. Experimental results show that as the fidelity of the predictive data provided by GNNs increases, the MAE and standard deviation of MLPSets-CHGNet on small experimental datasets decrease. Notably, models based on DenseGNN-CHGNet exhibit the highest accuracy, closely approximating the precision of DFT-calculated results. Similar conclusions can be drawn using the band gap dataset, as shown in Supplementary Figure S20.

Under high-fidelity predictive data, the error bars and standard deviation ranges of MLPSets-CHGNet on the experimental dataset decrease with an increase in training data volume, indicating that both high-fidelity predictive data and increased data volume are crucial for enhancing the model's transfer learning capability. Conversely, low-fidelity predictive data, such as those provided by GraphSAGE, do not significantly improve the transfer error and standard deviation on the experimental dataset even with increased data volume. This underscores the importance of data quality and highlights the critical role of comprehensive crystal structure graph representation strategies, which determine the quality of predictive data. The superior performance of DenseGNN-CHGNet confirms the effectiveness of the data-driven crystal structure graph representation strategy CHGNet-V1 proposed in this study, which provides predictive results close to DFT precision, demonstrating its high value as a knowledge transfer tool.

In Supplementary Figures S21a and c, we investigated the effect of transferring coGN and DenseGNN-CHGNet models trained on DFT formation energy datasets of different scales to the MLPSets-CHGNet small-scale experimental dataset. Our findings indicate that an increase in training data volume helps reduce the MAE and standard deviation of MLPSets-CHGNet on the experimental dataset. However, given the scarcity of DFT calculation data for many material properties, typically only about 1500 data points, as indicated by the red dashed line in the figures, we further tested the coGN-1500 and DenseGNN-CHGNet-1500 models trained on 1500 DFT formation energy data points (Figures S21b and d). These models were used to predict the formation energies of a large dataset comprising 53,000 structures, serving as training data. We then transferred the first-layer graph convolution (GC1) features from these trained models to the small-scale experimental dataset of MLPSets-CHGNet. Experimental results show that using the training data provided by the DenseGNN-CHGNet-1500 model improves the transfer MAE and standard deviation of MLPSets-CHGNet on the experimental dataset, whereas the use of training data from the coGN-1500 model does not significantly improve the model's performance. This suggests that DenseGNN-CHGNet exhibits significantly higher learning efficiency in transfer learning scenarios, effectively capturing physical laws similar to those obtained from large-scale, high-fidelity DFT data training, even with only 1500 DFT data labels. This finding emphasizes the importance of effective crystal structure graph representation strategies in enhancing model learning efficiency and provides a new approach to address the issue of data scarcity in materials science.

# Improving band gap prediction performance of disordered materials based on CHGNet-V1's structure graph representation

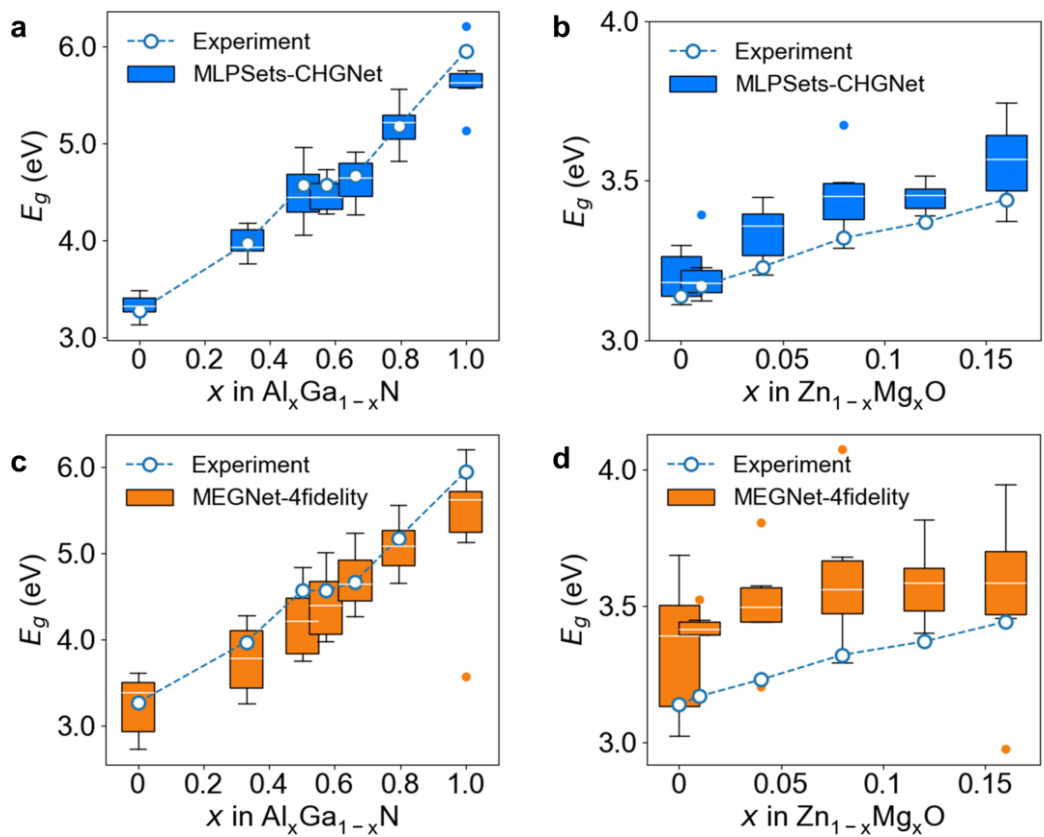

Figure 16 consists of four subplots, where (a) and (b) show the prediction results of the DenseGNN-CHGNet model after pretraining on the band gap dataset of Matbench, transferring to MLPSets-CHGNet, and fine-tuning on an ordered band gap experimental dataset. Subplots (c) and (d) present the prediction results of the MEGNet-4fi model. The error bars in each subplot represent one standard deviation. These results reflect the changes in the band gap (Eg) of $Al_xGa_{1-x}N$ and $Mg_xZn_{1-x}O$ materials as the composition variable $x$ varies.

For the MEGNet-4fi[49] model, it is trained by integrating PBE[50], GLLB-SC[51-53], HSE[54], and experimental data[55]. This multi-fidelity data fusion approach enables the model to achieve high accuracy in predicting the band gaps of disordered materials, reproducing the qualitative trends for all instances and achieving near-quantitative accuracy for most systems. Traditional machine learning models struggle to represent disordered material sites as linear combinations of atomic feature vectors, whereas graph network models like MEGNet-4fi can naturally represent disordered sites through element embeddings, which is a significant advantage. In contrast, the DenseGNN-CHGNet model benefits from the effectiveness of the data-driven crystal structure graph representation strategy, CHGNet-V1, proposed in this work. The model is first trained on the band gap dataset of Matbench, then transferred to the MLPSets-CHGNet framework, and further fine-tuned using 2645 ordered experimental data points. This process significantly reduces the MAE when predicting the band gaps of disordered materials like $Al_xGa_{1-x}N$[56] and

$Mg_xZn_{1-x}O^{57}$, demonstrating that the CHGNet-V1 strategy helps the model efficiently learn the physical laws underlying similar band gaps, enabling accurate predictions for complex materials.

In summary, the DenseGNN-CHGNet model, based on the data-driven crystal structure graph representation strategy CHGNet-V1, has made significant progress in predicting the band gaps of disordered materials, providing a new direction for developing band gap prediction models with near-experimental accuracy. This also highlights the respective advantages of different models in handling such problems: the MEGNet-4fi model excels due to its multi-fidelity data fusion, while the DenseGNN-CHGNet model achieves good results through the effective structure graph representation strategy.

## Discussion

Despite significant advancements in the application of GNNs in materials science, there are still notable shortcomings in representing electronic interactions. Existing models lack a structural graph representation strategy analogous to the "pseudopotential" concept in DFT, which is essential for effectively representing electronic interactions. Additionally, there is a lack of systematic comparative studies on the effectiveness of different crystal structure graph representation strategies. To address these issues, this paper systematically compares physics-based and data-driven crystal structure graph representation strategies based on the Streamlined MLPSets framework, ultimately confirming the superiority of the CHGNet-V1 and V2 representation strategies.

Through a comparative study of physics-based and data-driven crystal structure graph representation strategies, we have revealed their respective strengths and limitations. Physics-based strategies, such as MLPSets-AGNIFingerprints, MLPSets-OPSiteFingerprint, and MLPSets-CrystalNNFingerprint, excel in representing the local geometric topology of crystal structures but are limited in describing electronic interactions, symmetry, and long-range interactions. In contrast, advanced data-driven GNN strategies, such as MLPSets-M3GNET and the MLPSets-CHGNet series, perform more effectively in multi-dimensional representation of crystal structures. Notably, the MLPSets-CHGNet-V1 and V2 models utilize VAE to compress KS wave functions into a low-dimensional latent space using an encoder and then reconstruct them using a decoder, efficiently representing electronic structures. By combining this with the multi-body interaction modules of CHGNet and label feedback from the training dataset, these models achieve a comprehensive representation of crystal structures, outperforming existing advanced models in multiple benchmark tests. Specifically, MLPSets-CHGNet-V1 and V2 not only demonstrate superior performance in prediction accuracy, convergence speed, and learning efficiency on small datasets, but also exhibit excellent stability and accuracy in various material property extrapolation tasks. Even with limited data, MLPSets-CHGNet-V1 maintains low MAE, showcasing its strong learning efficiency and generalization capabilities. These results highlight the significant potential of data-driven graph representation strategies in enhancing model learning efficiency and extrapolation performance.

Based on the proposed advanced crystal structure graph representation strategy CHGNet-V1, we validate its effectiveness in three specific applications. First, extensive testing on 35 different benchmark datasets from Matbench and JARVIS-DFT shows that DenseGNN-CHGNet significantly outperforms existing state-of-the-art models across multiple datasets. Second, evaluation results indicate that the prediction data provided by DenseGNN-CHGNet is closest to DFT accuracy, and as the fidelity of pre-training data increases, the performance of MLPSets-CHGNet on experimental datasets also improves. Finally, by pre-training DenseGNN-CHGNet on the Matbench bandgap dataset and fine-tuning it with disorder experimental data, we significantly reduce the MAE in predicting the bandgap of disordered materials such as $Al_xGa_{1-x}N$ and $Mg_xZn_{1-x}O$. This demonstrates that the CHGNet-V1 crystal structure graph representation strategy helps the model efficiently learn the physical laws related to bandgap, enabling accurate predictions of complex material properties.

This research not only deepens the theoretical understanding of crystal structure graph representation but also provides important guidance for practical applications. By systematically comparing different graph representation strategies, we reveal the advantages of data-driven crystal structure graph representation strategies in material property prediction, especially in handling complex electronic interactions and extrapolation tasks. These findings offer new insights and technical approaches for developing efficient material prediction models, potentially accelerating the discovery and optimization of new materials. Future research can consider incorporating more physical and chemical prior knowledge, such as the "pseudopotential" concept, to further enhance the representation of electronic interactions in crystal structures. Additionally, exploring how to better handle complex many-body effects and long-range interactions in crystal structures can further improve model performance.

## Methods

**Implementation details of physics-based and data-driven crystal structure graph representation strategies**

AGNIFingerprints is employed to compute atomic fingerprints and direction-resolved fingerprints for crystal structures. These fingerprints are based on the integration of the product of the Radial Distribution Function (RDF) and a Gaussian window function. Specifically, the process begins by identifying the neighboring atoms around a given atom and calculating the distances between them. A cutoff function is then applied to limit the range of distances considered. For a given set of width parameters (etas), the product of each distance with its corresponding Gaussian weight, multiplied by the cutoff function, is computed. If direction-resolved fingerprints are required, projections in various directions are further calculated. The final fingerprint values for each parameter are obtained by summing these products and combining them into the ultimate output.

OPSiteFingerprint is used to calculate the Local Structure Order Parameters (LSOPs) at specific sites within a crystal structure. LSOPs are determined based on the characteristics of the neighboring environment around a given site. In the computation, the neighboring atoms around an atom are first identified, and the corresponding neighbor shells are found according to the expected coordination number (CN). For example, for tetrahedral LSOPs, the four nearest neighbors are identified; for octahedral LSOPs, the six nearest neighbors are identified; and for body-centered cubic (BCC) LSOPs, the eight nearest neighbors are identified. If such shells cannot be found, the actual observed coordination numbers are evaluated. To ensure robustness against numerical variations, the concept of relative distances is introduced, and neighbors are binned accordingly. Additionally, multiple width variables (dr, ddr, ndr) are introduced to test LSOP values at different widths. For each width variable, the calculated LSOP values are stored in a dictionary, and peak positions are determined through histogram analysis. Finally, by combining the LSOP values across all width variables, a fingerprint vector that characterizes the local geometric structural features of the site is generated. These features reflect the local geometric structural environment surrounding the site.

CrystalNNFingerprint is utilized to compute the local order parameter fingerprints at specific sites in a periodic crystal. This fingerprint represents the values of various order parameters at that site. The "wt" order parameter describes the consistency of the site with a specific coordination number, while the other order parameters are derived by multiplying the "wt" with the corresponding order parameter values. The computation starts by using the CrystalNN method to obtain the nearest neighbor data for a given site. Then, the "wt" values for each coordination number are determined based on the expected CN, and the corresponding order parameters are calculated. If chemical information (such as electronegativity differences) is provided, chemically related features are also computed. Specifically, for each coordination number, the "wt" value is calculated, and the weighted order parameters are computed using the "wt" value and the corresponding order parameter values. If chemical information is available, the differences in chemical properties are calculated and incorporated into the fingerprint vector. The fingerprint vector includes the weighted order parameters and chemical property differences, providing a comprehensive description of the local structure and chemical environment around the site. These features help in understanding and characterizing different dimensions of the crystal structure, thus offering detailed insights into the local properties of the crystal structure.

The three physics-based calculators (AGNIFingerprints, OPSiteFingerprint, and CrystalNNFingerprint) cover five key aspects in the representation of crystal structures:

1. Chemical information integration: AGNIFingerprints and OPSiteFingerprint do not explicitly incorporate information from the periodic table or pre-trained knowledge (which we introduce in this work). In contrast, CrystalNNFingerprint can embed chemical information, such as electronegativity and ionization energy, through the chem_info parameter, but this requires manual configuration.
2. Geometric topological information: All three methods represent the geometric topological information of crystal structures. AGNIFingerprints and OPSiteFingerprint describe the local geometric environment by calculating neighbor distances and distributions, while CrystalNNFingerprint uses nearest neighbor data and weight information to describe the geometric structure.
3. Electronic interaction information: AGNIFingerprints and OPSiteFingerprint primarily focus on geometric information and do not explicitly represent electronic interaction information. CrystalNNFingerprint, however, can embed some electronic interaction information through the chem_info parameter.
4. Symmetry information: None of the three methods explicitly address symmetry information. They mainly focus on local geometric and electronic properties rather than overall symmetry.
5. Long-Range interaction information: AGNIFingerprints and OPSiteFingerprint are primarily localized and do not explicitly capture long-range interaction information.

Overall, these three methods excel in representing the local geometric and electronic properties of crystal structures but fall short in capturing symmetry and long-range interaction information.

The data-driven GNN-based graph representation strategies were trained on the Materials Project Trajectory Dataset, which contains approximately 1.58 million atomic configurations and dynamic evolution information for 145,000 inorganic materials. The combination of large-scale training datasets with advanced GNN model designs enhances the representation of geometric, electronic interactions, symmetry, and long-range interaction information in crystals. For example, the M3GNET model introduces a three-body interaction module to accurately describe geometric topology and employs rotation-invariant and symmetry-preserving operations to capture crystal symmetry. It also incorporates global state information (such as charge, temperature, and pressure) and message-passing mechanisms to represent multi-scale crystal structures. CHGNet adopts an innovative graph neural network architecture and many-body interaction modeling to comprehensively represent key information in crystal structures. It converts atomic numbers into high-dimensional feature vectors using element embeddings to capture elemental properties. CHGNet aggregates geometric features, such as distances and angles between neighboring atoms, using crystal graphs and auxiliary bond graphs. Multi-layer graph convolutions expand the range of information propagation, capturing large-scale geometric structures. Additionally, CHGNet incorporates magnetic moment information as a charge state constraint and integrates electronic structure information generated by a VAE model based on KS wave functions, enabling the representation of electronic interactions. The graph convolution layers naturally handle permutation symmetry and indirectly reflect macroscopic symmetry characteristics. By calculating three-body interactions and performing multi-layer graph convolutions, CHGNet effectively captures long-range interaction features.

**Dataset and training details for the VAE Model**

In the process of generating electronic structure features, this study references the VAE method proposed by Bowen Hou. Based on the Computational 2D Materials Database (C2DB), a carefully designed dataset comprising 302 diverse samples was created, covering a wide range of materials, including metals and semiconductors, with each material's unit cell containing 3 to 4 atoms. The DFT electronic states of these materials were sampled on a uniform 6×6×1 k-point grid for unsupervised training of the VAE. The dataset includes 68,384 DFT electronic states, which are randomly divided into a 90% training set and a 10% testing set.

The VAE model architecture consists of an encoder and a decoder. The encoder maps the high-dimensional KS wave function magnitudes to a low-dimensional latent space vector, which contains variational means and variances. The decoder is responsible for reconstructing the input wave functions from this low-dimensional latent vector. During training, the VAE optimizes its parameters by minimizing the mean squared error (MSE) between the reconstructed wave functions and the original wave functions, as well as the Kullback-Leibler (KL) divergence between the latent space distribution and the standard normal distribution. The introduction of the

KL divergence ensures the smoothness and normality of the latent space, which is crucial for the model's generalization and generation capabilities.

To enhance the model's representation of crystal symmetry, we incorporated a Global Average Pooling (GAP) layer in the encoder to handle inputs of different sizes and produce fixed-length outputs. Additionally, we used cyclic padding techniques to account for the periodic boundary conditions of the crystal systems. Furthermore, the first layer of the encoder employs a discrete rotation convolutional neural network (CNN) to maintain invariance to lattice vectors and directional choices. These design choices enable the VAE to capture properties related to the physical symmetries of crystals without explicitly including symmetry labels. Through this approach, the VAE model not only effectively compresses the high-dimensional KS wave function magnitudes into a low-dimensional latent space but also accurately reconstructs the original wave functions during decoding, preserving key physical information. This training process allows the VAE to autonomously determine the optimal representation of the electronic structure.

**MLPSets models framework**

The MLPSets model framework is designed to minimize the interference of GNN operations, such as message passing and graph convolution, by simplifying the network design, thereby enabling a clear comparison of different crystal structure graph representation strategies. The framework accepts various graph representation inputs, including edge indices, atomic numbers, offsets, and atomic fingerprints, to construct a graph where atomic numbers and other attributes (such as atomic fingerprints) serve as node features, and Euclidean distances between atoms serve as edge features. In MLPSets, the depth parameter is set to 0, meaning no graph convolution operations are performed, ensuring the simplicity of the model design. This approach allows the MLPSets framework to focus on evaluating the effectiveness of different graph representation strategies without the influence of complex GNN architectures. Additionally, the framework employs a five-fold cross-validation method for model training and evaluation to ensure the accuracy and reliability of the results. Through this series of design choices, the MLPSets framework not only effectively compares physics-based and data-driven crystal structure graph representation strategies but also identifies the optimal representation methods, providing strong support for material property prediction.

**Data and model training**

The input for the MLPSets includes atomic offset positions, atomic numbers, atomic fingerprint vectors (from physics-based calculators and data-driven features representation), and edge indices describing atomic connections. In terms of training strategy, all models use a batch size of 256, a total of 300 epochs, and validation checks every 20 epochs. Training employs an Adam optimizer with an exponential decay strategy, starting with an initial learning rate of 0.001 and decaying by a factor of 0.5 every 5800 steps. The loss function selected is the MAE. Model evaluation uses 5-fold cross-validation with a random seed of 42 and data shuffling. Prior to training, input data are standardized to ensure a mean of 0 and a standard deviation of 1. For each model training, the data are randomly split into 80%-10%-10% train-validation-test sets, repeated five times with random seeds ranging from 0 to 4. Training is terminated early if the MAE on the validation set does not improve for 300 consecutive epochs, selecting the model with the lowest validation error as the best model. Each model is trained five times, reporting the mean and standard deviation of the metrics on the test set. To facilitate direct comparisons with existing models, we use the same 5-fold shuffle splitting/stratified splitting and random seeds as those used in Matbench and Jarvis-DFT. In this setting, each 80% of the training data is further split into 90%-10% train-validation sets, and the validation set is used in the same manner as in other fittings.

For DenseGNN-CHGNet, we employ a KNN (K=12) method for edge selection to optimize performance. Considering variations in edge length due to changes in crystal density, we apply a Gaussian kernel function to extend edge lengths and use them as edge features. MAE serves as the standard evaluation metric for material property prediction tasks.

## Datasets descriptions

The JARVIS-DFT dataset was developed using the Vienna ab initio simulation package (VASP). Most properties are calculated using the OptB88vdW functional[58]. For a subset of the data, we use TBmBJ[59] to get a better bandgap. We use density functional perturbation theory (DFPT)[60] to predict piezoelectric and dielectric constants with electronic and ionic contributions. The linear response theory-based frequency[61] based dielectric function was calculated using OptB88vdW and TBmBJ, and the zero energy values are used to train ML models. The TBmBJ frequency-dependent dielectric function is used to calculate the maximum efficiency limited by the spectrum (SLME)[62]. The magnetic moment is calculated using spin-polarized calculations considering only ferromagnetic initial configurations and ignoring any DFT + U effects. Thermoelectric coefficients such as the seebeck coefficient and power factor are calculated using the BoltzTrap software[63] with a constant relaxation time approximation. The exfoliation energy of van der Waals bonded two-dimensional materials is calculated by calculating the energy difference between each atom in the bulk phase and the corresponding monolayer. Spin orbit spillage[64] is computed as the disparity between material wavefunctions with and without the inclusion of spin-orbit coupling effects. All JARVIS-DFT data and classical force field inspired descriptors (CFID)[65] are generated using the JARVIS-Tools software package. The CFID baseline model is trained using the LightGBM software package[65].

Matbench is an automated benchmark testing platform specifically designed for the field of materials science, designed to evaluate and compare the most advanced ML algorithms that predict various solid material properties. It provides 13 carefully curated ML tasks that cover a wide range of inorganic materials science, including the prediction of various material properties such as electronics, thermodynamics, mechanics to thermal properties of crystals, two-dimensional materials, disordered metals, etc. The datasets for these tasks come from different density functional theories and experimental data, with sample sizes ranging from 312 to 132,000. The platform is hosted and maintained by the MP, providing a standardized evaluation benchmark for the field of materials science.

The dataset for liquid and amorphous structures comprises 2000 silicon (Si) structures, with 50% being liquid and 50% being amorphous. Data for different dimensions (0D (Clusters), 2D (Sheets), and 3D (Bulk)) are sourced from the Samara Carbon Allotrope Database (SACADA). This database provides a dataset of 511 block structures for multi-scale material classification tasks. The 2D structure data is obtained through the CASTING framework and LCBOP potential sampling of carbon samples. For 0D structures, a dataset of 704 carbon nanoclusters is utilized. The crystal structure dataset based on space groups consists of 10,517 crystal structures, covering 7 crystal categories of 8 different space groups. This dataset is sourced from the Pymatgen software package[66], a Python library for materials analysis capable of processing and analyzing crystal structure data.

# Acknowledgements

We are grateful for the financial support from the National Key Research and Development Program of China (Grant Nos.2021YFB3702104). The computations in this paper were run on the π 2.0 cluster supported by the Center for High Performance Computing at Shanghai Jiao Tong University.

# Contributions

H.D and H.W devised the idea for the paper. H.D implemented the idea and conducted the code design and visualizations. H.D and H.W interpreted the results and prepared the manuscript.

# Code Availability

The data that support the findings of this study will be uploaded to https://github.com/dhw059/RSE in the future. The source code of DensGNN is freely accessible at https://github.com/dhw059/DenseGNN. The dataset used for space group classification is available via https://www.nomad-coe.eu/49. The Carbon bulk structures used in this work are available via SACADA database https://www.sacada.info/. All the other datasets used for different classification task are available at https://github.com/sbanik2/CEGANN.